\documentclass[aip,reprint,twocolumn,superscriptaddress,floatfix,showpacs,nofootinbib]{revtex4-1}
\usepackage{bm}
\usepackage{textcomp} 
\usepackage{enumitem}
\usepackage[svgnames]{xcolor}
\usepackage{dcolumn}     
\usepackage[english]{babel} 
\usepackage[thicklines]{cancel}
\usepackage{times} 
\usepackage{tcolorbox}
\usepackage{xfrac}
\usepackage{graphicx} 
\graphicspath{{figures/}} 
\usepackage{booktabs} 
\usepackage{amsfonts,mathrsfs,amsmath,amssymb,mathtools} 
\usepackage{wrapfig}
\usepackage{empheq}
\usepackage{float}
\usepackage{tikz}
\usepackage{tikz-3dplot}
\usepackage{braket}
\usepackage[pdftex,colorlinks=true,urlcolor=blue,linkcolor=blue,citecolor=violet]{hyperref}
\usepackage[font=small,labelfont={sf,bf},justification=raggedleft,justification=raggedright]{caption}
\usepackage{subcaption}
\usepackage{threeparttable}
\usepackage{accents}
\usepackage{booktabs}
\usepackage{multirow}
\usepackage{chngcntr}
\usepackage{pgfplots}
\usepackage{cleveref}
\usepackage{verbatim}
\usepackage{listings}
\usepackage{ulem}

\usetikzlibrary{external,arrows,patterns,decorations.pathmorphing,intersections}
\newcommand{\fracinline}[2]{\frac{\raisebox{-0.4ex}{$#1$}}{\raisebox{-0.1ex}{$#2$}}}
\newcommand{\fracinlines}[2]{\frac{\raisebox{-0.4ex}{$#1$}}{\raisebox{-0.3ex}{$#2$}}}
\newcommand{\fracinliness}[2]{\frac{\raisebox{-0.3ex}{$#1$}}{\raisebox{+0.3ex}{$#2$}}}

\newcommand*{\complexconjugate}[1]{\ensuremath{{#1}^*}}
\newcommand\scalemath[2]{\scalebox{#1}{\mbox{\ensuremath{\displaystyle #2}}}}
\newcommand{\vnucl}[1]{V_{\text{ext}}(#1)}

\renewcommand{\thefigure}{\arabic{figure}}

\newcommand{\cl}[1]{{\color{green} #1}}   
\newcommand{\ecorr}[0]{\mathcal{E}_{\text{corr}}}

\begin{document}

\title{Production of positronium chloride: A study of the charge exchange reaction between Ps and Cl$^{-}$}
\author{K. Lévêque-Simon}
\affiliation{Laboratoire de Chimie Physique-Matière et Rayonnement, Sorbonne Université and CNRS, F-75005 Paris, France}
\author{A. Camper}
\affiliation{Department of Physics, University of Oslo, Sem Saelandsvei 24, 0371 Oslo, Norway}
\author{R. Ta\"ieb}
\affiliation{Laboratoire de Chimie Physique-Matière et Rayonnement, Sorbonne Université and CNRS, F-75005 Paris, France}
 \author{J. Caillat} 
 \affiliation{Laboratoire de Chimie Physique-Matière et Rayonnement, Sorbonne Université and CNRS, F-75005 Paris, France}
 \author{C. Lévêque}
 \affiliation{Laboratoire de Chimie Physique-Matière et Rayonnement, Sorbonne Université and CNRS, F-75005 Paris, France}
\email{kevin.leveque-simon@sorbonne-universite.fr}
\author{E. Giner}
\affiliation{Laboratoire de Chimie Théorique, Sorbonne Université and CNRS, F-75005 Paris, France}
\date{\today}
\begin{abstract}
{\small We present cross sections for the formation of positronium chloride (PsCl) in its ground state from the charge exchange between positronium (Ps) and chloride (Cl$^-$) in the range of 10 meV\,-\,100 eV Ps energy. We have used theoretical models based on the first Born approximation in its three-body formulation. We simulated the collisions between Ps and Cl$^-$ using \textit{ab-initio} binding energies and positronic wave functions at both mean-field and correlated levels extrapolated to the complete basis set limit. The accuracy of these \textit{ab-initio} data was benchmarked on the PsF system with existing highly accurate results including very recent quantum Monte Carlo results. We have investigated Ps excited states up to ${n=4}$. The results suggest that the channel Ps(${n=2}$) is of particular interest for the production of PsCl in the ground state, and shows that an accurate treatment of the electronic correlation leads to a significant change in the global shape of the PsCl production cross section with respect to the mean-field level.}
\end{abstract}

\maketitle

\section{Introduction}
Positronium Chloride (PsCl) is a compound made of matter and anti-matter (one positron), the existence of which was first predicted in 1953 \cite{Simons_1953}. A first experimental evidence was obtained a decade later, in 1965, through the observation of a ``shoulder" in positron annihilation lifetime spectra, interpreted as the production of PsCl by the capture of one positron in a mixture of chlorine and argon gas  \cite{Tao_1965}. The proposed mechanism was a charge exchange reaction between chlorine (Cl$_{2}$) molecules and positronium (Ps) atoms produced by collisions between positrons (e$^{+}$) and argon atoms. In spite of such significant insight, no direct measurement of the binding energy of PsCl had been reported while first computations of its energy spectrum were performed in 1992\cite{Schrader_1992}. To date, our knowledge on PsCl is still extremely sparse. However, the accumulation and storage of positrons in Surko-Penning-Malmberg traps \cite{Surko_2004} offer possibilities to form PsCl in\cl{,} so far\cl{,} unexplored ways and within a time-window of a few nanoseconds so that it can be synchronized with laser pulses for spectroscopy. This perspective has been identified in recent theoretical studies investigating the formation of positron binding atoms through laser-assisted photorecombination \cite{Surko_2012}, charge exchange reactions between positronium atoms excited in Rydberg levels \cite{Antonello_2020}, neutral atoms \cite{Swann_2016}. The AEgIS (Antihydrogen Experiment: gravity, Interferometry, Spectroscopy)\cite{Aegis_2010,Aegis_2019} project has demonstrated pulsed production of antihydrogen via a charge exchange reaction two years ago \cite{Amsler_2021}. 

Regarding now the simulation of positronic systems, there are two types of experimentally relevant quantities: the observables that can be fully computed using \textit{ab initio} techniques, and the more complex cross sections which necessarily imply further approximations. Regarding the first category, the crucial quantities to be determined are typically the energy binding of the positron to the electronic system, the annihilation rate of the electron-positron yielding to the lifetime of the positron and the normalization integral describing the spatial attachment of the positron to the electronic system. These quantities can be obtained using either many-body perturbation theory (MBPT) \cite{Pos-GW-GriLud-PRA-04,Pos-GW-LudGri-10,Pos-GW-HofCunRawPatGre-Nat-22,Pos-GW-RawHofCunPatGre-PRL-23}, or more standard wave function based \textit{ab initio} approaches such as Quantum Monte Carlo (QMC) \cite{Pos-QMC-SchYosIgu-PRL-91,Pos-QMC-BreMelMor-JCP-98,Pos-QMC-JiaSch-JCP-98,Pos-QMC-MelMorBre-JCP-99,Pos-QMC-Bor-APPA-05,Pos-VMC-KitMaeTacTowNee-JCP-09,Pos-VMC-DruLopNeePic-PRL-11,Pos-QMC-ChaJorBarTka-JCTC-22}, explicitly correlated gaussians\cite{BubAda-JCP-04} and post-Hartree-Fock (HF) calculations\cite{Saito_2003, Saito_2005}. A commonly acknowledged fact is that these quantities strongly depend on the level of treatment of the correlation effects arising between the electrons and the positron. An advantage of QMC approaches is certainly the possibility to use very complex parametrization for the variational wave functions thanks to the use of explicit electron-positron correlation factors. The use of correlation factors enables to satisfy the electron-positron cusp conditions and more generally to describe the short-range correlation effects, to the price of stochastic optimization techniques. 

On the other hand, post-HF and MBPT approaches can easily access the relevant spectroscopic quantities but require relatively high angular momentum expansions and further basis set extrapolation techniques in order to mitigate the basis set incompleteness error which mostly comes from the short-range correlation effects. Nevertheless, it has been shown\cite{Pos-GW-LudGri-10} that all relevant quantities do not convergence at the same rate with respect to the maximum angular momentum used in the basis. According to the study of Ludlow \textit{et al.}\cite{Pos-GW-LudGri-10}, the binding energy of the positron converges much faster than the annihilation rate with respect to the maximum angular momentum used in the basis set. The latter can be expected as the annihilation rate is a direct measure of the on-top electron-positron pair density averaged over the positronic system, while the binding energy largely benefits from the cancellation of errors between the description of the mixed positronic system and the purely electronic system. 

In the present manuscript, we investigate the mixed electronic/positronic structure of PsCl, followed by its formation through a charge exchange reaction in a similar way to the seminal GBAR (Gravitational Behaviour of Antihydrogen at Rest)\cite{GBAR_2011,GBAR_2019} production scheme for antihydrogen. 
We therefore investigate not only the binding energy of the positron but also the cross section involved in the charge exchange. 

When the computation of a cross section is required, further approximations than the usual Born-Oppenheimer approximations are mandatory. A commonly used strategy to simulate the scattering process is the so-called first Born approximation (FBA)\cite{Swann_2016} which in our case requires the following ingredients: i) the positron binding energy of the positronium chloride system, ii) the single-particle picture (SPP) of the positron wave function embedded in the electronic system, and iii) the one-electron densities of PsCl and Cl$^-$, where the latter is needed to determine the effective screened Coulomb potentials involved in the continuum part of the collision process. Therefore, the computation of cross sections does not require to evaluate the electron-positron contact density\cite{Pos-GW-HofCunRawPatGre-Nat-22} linked to the annihilation rates, which very slowly converge with the quality of the basis set in post-HF and MBPT approaches. However, accurate values of the energy differences are required, especially between PsCl and Cl$^-$ given ii).

In order to investigate the dependency of quantities related to the collision process with respect to the level of calculations, we use post-HF \textit{ab-initio} methods from quantum chemistry. The use of restricted Hartree-Fock (RHF), configuration interaction with single and double substitutions (CISD), and linearized coupled cluster with single and double excitations\cite{Bartlet-ARPC-81} (LCCSD) in increasingly large basis sets, together with basis set extrapolation techniques \cite{HelKloKocNog-JCP-97,HalHelJorKloKocOls-CPL-98}, allows us to investigate the impact of correlation effects on the cross section of PsCl formation. We first benchmark our methodology by computing the electron binding energy of an other positronium halide, namely PsF, and compare our results with the state-of-the-art QMC calculations together with multi reference CI (MRCI) and MBPT calculations in B-spline basis sets extrapolated to the complete basis set (CBS) limit. We show that the positron binding energy obtained by extrapolating our LCCSD results agrees within 1 mH with both the MBPT results of Ref. \onlinecite{Pos-GW-LudGri-10} and with the diffusion Monte Carlo (DMC) using an antisymmetrized geminal power (AGP) trial wave function of Ref. \onlinecite{Pos-QMC-ChaJorBarTka-JCTC-22}. We continue with the computation of the binding energy of PsCl and compare with the available MRCI and MBPT results. We then focus on the simulation of the cross section of PsCl, and show that the correlation treatment affects both qualitatively and quantitatively the latter by significantly changing both its shape and magnitude. We further assess the validity of our results through a systematic investigation of other collision related quantities with respect to the quality of the basis. We observe that the main dependence  of the PsCl cross section to the basis set is induced by the slow convergence of the energy differences corresponding to the attachment/detachment of particles (either electron or positron) in correlated treatment. 

The paper is organized as follows. Details on our theoretical models are described in section \ref{section-1}. The results are  discussed in section \ref{section-2}. We conclude and discuss some perspectives of our work in section \ref{section-3}. Atomic units are used throughout the paper unless specified otherwise.

\begin{figure}[bp!]
\centering
\includegraphics[scale=1.05]{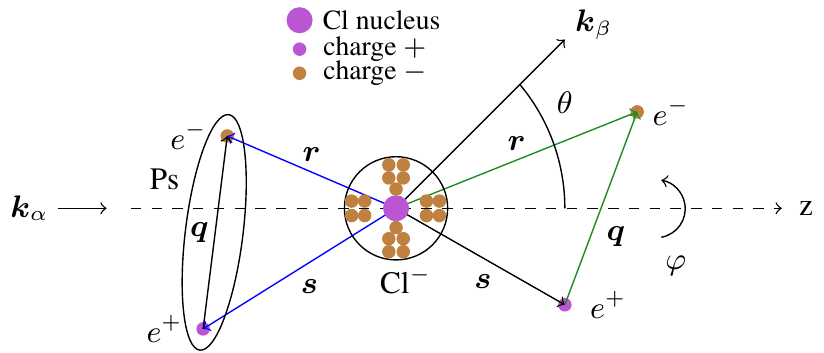}
\caption{Coordinates used for the studied charge exchange reaction given in Eq. \eqref{eqn:the_reaction}. The bound states present in the entrance channel are circled. The blue and green straight lines correspond to the coordinates which are used to define the effective short-range perturbative potential in \textit{prior} and \textit{post} form, respectively. The wavevectors $\boldsymbol{k}_{\alpha}$ and $\boldsymbol{k}_{\beta}$ are those of the incident Ps and the ejected electron, such that the diffusion angle is defined by ${\theta=(\hat{\boldsymbol{k}}_{\alpha},\hat{\boldsymbol{k}}_{\beta})}$. The direction chosen for ${\boldsymbol{k}_{\alpha}}$ implies that the collision is invariant by a rotation of angle ${\varphi}$ around the z-axis.\label{Figure-1}}
\end{figure}

\section{Methodology\label{section-1}}

In this section, we first present the general approach used to model the collision reaction of interest and compute the related cross sections. We then provide technical details on the atomic structure computations to obtain energies, electronic densities and single particle wavefunctions needed as input for the cross section evaluation. This part, which consists in adapting standard quantum chemistry approaches to account for the presence of a positron in the electronic cloud of chemical species, represent the main methodological development of the present study.

\subsection{First Born approximation models for the treatment of charge exchange between Ps and Cl$^-$}

Cross sections related to the production of PsCl in the ground state are obtained using perturbative approaches based on the first Born approximation. We will consider the reaction
\begin{eqnarray}\label{eqn:the_reaction}
{\text{Ps}(n\ell)+\text{Cl}^-\!\rightarrow\text{PsCl}(1s\scalemath{0.8}{+})+e^-},
\end{eqnarray}
allowing to extend an existing open-source collision code\footnote{See \url{https://gitlab.com/k.levequesimon/thesis}.} initially designed for the GBAR experiment to more complex cases, which deals with the following three-body charge exchange processes\cite{Leveque_2020,Comini_2013,Comini_2021}: ${\text{Ps}(n\ell)+\overline{p}\!\rightarrow\overline{\mathrm{H}}(1s)+e^-}$. In this reaction, $\overline{p}$ stands for antiproton and $(n\ell)$ for the quantum numbers characterizing the positronium state, as Ps can be prepared in an excited state. The quantum treatment of such many-body collision reactions requires to describe both the continuum and bound parts of the collision partners in the entrance and exit channels. For the continuum part, one may include Coulomb distortions - Continuum Distorted Wave Initial State (CDW) - or discard them - Coulomb Born Approximation (CBA) - in the entrance channel. These distortions occur when the electron and positron that forms the incident Ps are in the vinicity of the antiproton target. In the exit channel, the propagation of the ejected electron is described by a plane wave due to the neutral charge of the produced antihydrogen atom. 

Adapting the first reaction of GBAR to the production of PsCl by substituting the antiproton that captures a positron by a chlorine anion requires, first, to compute the structural data associated with PsCl. These input data correspond to the one-electron density of PsCl (and Cl$^-$ also if the CDW formalism is employed), the single-particle picture of the ground state radial wavefunction of the positron bound to Cl$^-$, and the corresponding positron binding energy. The $S$-matrix associated with the collision is computed in the \textit{post} form, which means that the chosen effective short-range perturbative potential describes the propagation of the ejected electron in the continuum of the produced positronium chloride. In addition, the many-body bound parts of this perturbative potential can be described using standard quantum chemistry calculations, adapted here for the inclusion of a positron in the electronic structure of Cl$^-$ as detailed in subsection~\ref{sec:QCwithe+}. Using the spatial coordinates depicted in Fig. \ref{Figure-1}, the latter is expressed as
\begin{equation}
V_{\beta}(\boldsymbol{r},\boldsymbol{s})=\fracinlines{1}{|\boldsymbol{r}|}-\fracinlines{1}{|\boldsymbol{q}|}+\bigg[\int_{\mathbb{R}^3}\!\!\!d\boldsymbol{r}'\fracinlines{\rho_{\beta}(\boldsymbol{r}')}{|\boldsymbol{r}-\boldsymbol{r}'|}-\fracinlines{(Z_A+1)}{|\boldsymbol{r}|}\bigg].\label{es.1}
\end{equation}
In the above equation, ${\rho_{\beta}(\boldsymbol{r})=\sum_{i=1}^{Z_A}|\phi_i(\boldsymbol{r})|^2}$ is the electronic density of PsCl, $\boldsymbol{r}$ and $\boldsymbol{s}$ are respectively the coordinates of the electron and of the positron with respect to the chlorine nucleus with  atomic number ${Z_A=\text{17}}$. The relative coordinate ${\boldsymbol{q}=\boldsymbol{r}-\boldsymbol{s}}$ coincides with the Ps center of mass in the entrance channel. In the framework of GBAR (${\text{Ps}+\overline{p}}$), the square brackets term in Eq. (\ref{es.1}) vanishes as the antiproton charge is equal to $-$1 and that no electron constitutes ${\overline{\mathrm{H}}}$. As a result, the one-electron density ${\rho_{\beta}}$ of the produced positronic atom is zero. In the present framework (${\text{Ps}+\text{Cl}^-}$), the term ${-Z_A/|\boldsymbol{r}|}$ in square brackets describes the Coulomb attractive interaction between the ejected electron and the nucleus. In counterpart, the Hartree potential allows the presence of more than three bodies in the charge exchange processes, by taking into account that the ejected electron feels, at a distance $r$ from the nucleus, an electronic charge density\cite{Hervieux_2019}. Details of the cross section calculations are given in Appendix \ref{appendix-1}. 

\subsection{\textit{Ab-initio} treatment of a mixed positronic/electronic system}\label{sec:QCwithe+}

In the framework of the Born-Oppenheimer and infinite mass approximations, the Hamiltonian of a chemical species with a positron embedded in its $N$-electron cloud is given by
\begin{align}
&\mathcal{H}=\sum_{i=1}^N\Big(-\fracinline{1}{2}\Delta_i+\vnucl{\boldsymbol{r}_i}\Big)+\sum_{i=1}^{N}\sum_{j>i}\fracinline{1}{|\boldsymbol{r}_i-\boldsymbol{r}_j|}\nonumber\\[-1mm]
&\phantom{\mathcal{H}=}-\fracinline{1}{2}\Delta_p-\vnucl{\boldsymbol{r}_p}
-\sum_{i=1}^N\fracinline{1}{|\boldsymbol{r}_i-\boldsymbol{r}_p|}\label{es.2}.
\end{align}
In this expression, ${\vnucl{\boldsymbol{r}}=-\sum_{A}Z_A/|\boldsymbol{r}-\boldsymbol{R}_A|}$ is the external electrostatic potential created by the nuclei of charge ${Z_A}$ located at ${\boldsymbol{R}_A}$. The spatial coordinates $\boldsymbol{r}_i$ and $\boldsymbol{r}_p$ are those of the $i$-th electron and the positron, respectively. The two first terms correspond to the purely electronic Hamiltonian, the next two to the Hamiltonian of the positron, and the last one to the Coulomb (attractive) interactions between the electrons and the positron.

With respect to the usual purely electronic Hamiltonian, one needs to consider an additional particle of positive charge 
interacting with the $N$ electrons and the nucleus
\footnote{Since we consider a single positron in absence of an external magnetic field, its spin plays no role and is discarded in the present study.}. 
Our electronic and positronic structure calculations rely on three levels of approximations: the RHF, CISD and LCCSD approaches 
and we give hereafter the main differences with respect to their usual  purely electronic versions. 
The SPPs used here are either the RHF orbitals or the natural orbitals of the LCCSD wave function projected 
on the CISD subspace (\textit{i.e.} the RHF determinant and the single- and double substitutions), 
both for electrons and for the positron. 

The RHF approach needs to be extended in two ways: i) one needs to optimize the orbitals of the positron and ii) the orbitals of the electrons have to be optimized taking into account the presence of the positron. Point i) is achieved by expanding the positron's orbitals on a basis of atomic orbitals (AOs) set and computing the corresponding Fock matrix which accounts for the purely one-body part together with the attractive Coulomb interaction with the electrons. Point ii) is dealt with by adding the attractive Coulomb interaction with the positron's charge density to the usual Fock matrix. In our framework, we use the same AOs basis set for both electrons and positron for practical reasons. The equations related to the Fock matrices for both electrons and the positron are given in Ref. \onlinecite{Kurtz_1980}. As usual in Roothan-Hall equations, a damping of the order of 20$\%$ is also introduced in the self-consistent process to avoid convergence problems. 

Regarding the CISD approach, the Slater-Condon rules are changed by the presence of the positron and the size of the CISD matrix to be diagonalized is larger. For CI methods in general, any element of the Hilbert space is a Slater determinant $\ket{I}$ which is expressed as ${\ket{I}=\ket{I_\alpha}\otimes\ket{I_\beta}\otimes\ket{i_p}}$, where ${\ket{I_{\alpha/\beta}}}$ are Slater determinants of $\alpha/\beta$ electrons, and $\ket{i_p}$ is the orbital occupied by the positron. Regarding now the composition of the CISD wavefunction, it contains additional terms with respect to the usual purely electronic wavefunction, which are: 
\begin{itemize}[noitemsep,nolistsep]
\item all usual single and double electronic excitations with the positron in the lowest orbital,
\item all positronic single excitations with the electrons described by the HF Slater determinant, 
\item all products of single excited electronic determinants by single excited positronic determinants. 
\end{itemize}

The LCCSD equations of Bartlet\cite{Bartlet-ARPC-81} are implemented as a self-consistent coupled electron pair zero 
(CEPA0) following the work of Alrichs\cite{Ahlrichs-CPC-79}. 
The latter corresponds in practice in an iterative dressing of the CISD matrix as follows: 
 \begin{itemize}[noitemsep,nolistsep] 
 \item at a given iteration $n$, one adds the correlation energy $\mathcal{E}_{\text{corr}}^{(n)}$ to all the diagonal element of the CISD matrix 
       except the Hartree-Fock Slater determinant, 
 \item the dressed CISD matrix is then diagonalized to obtain new CI coefficients and correlation energy $\mathcal{E}_{\text{corr}}^{(n+1)}$, 
 \item the iteration is stopped when the correlation energy is converged within to $10^{-6}$ au.
 \end{itemize}

\begin{widetext}
\begin{center}
\begin{table}[htb]
\begin{ruledtabular}
\begin{tabular}{|l|c c c| c c c| c c c|}
  Basis/System        &\multicolumn{3}{c|}{F}             &\multicolumn{3}{c|}{F$^-$}                & \multicolumn{3}{c|}{PsF}            \\ 
\hline
                &RHF       & $\ecorr$(CISD) &  $\ecorr$(LCCSD)  &   RHF       &  $\ecorr$(CISD) &  $\ecorr$(LCCSD)     &  RHF       & $\ecorr$(CISD) &  $\ecorr$(LCCSD) \\ 
  AVTZ-MOD      &-99.40218 &-0.21684        & -0.22550          &   -99.45083 & -0.27799        & -0.29626             &  -99.63445 & -0.30157       &  -0.33111  \\
  AVQZ-MOD      &-99.40925 &-0.23350        & -0.24286          &   -99.45747 & -0.29666        & -0.31604             &  -99.64140 & -0.32298       &  -0.35584  \\
  AV5Z-MOD      &-99.41120 &-0.23976        & -0.24934          &   -99.45926 & -0.30352        & -0.32241             &  -99.64322 & -0.33125       &  -0.36508  \\
Estimated CBS   &-99.41120 &-0.24632        & -0.25613          &   -99.45926 & -0.31071        & -0.32908             &  -99.64322 & -0.33993       &  -0.37479  \\
\hline
\hline
Energy difference &\multicolumn{3}{c|}{EA = E(F)-E(F$^-$)} &    \multicolumn{3}{c|}{PBE = E(F$^-$)-E(FPs)}   &\multicolumn{3}{c|}{BE = E(F) + E(Ps)$^a$ - E(FPs)} \\
\hline
                &  RHF     &  CISD    &  LCCSD      &    RHF       & CISD      & LCCSD          &   RHF      &  CISD    &  LCCSD    \\
  AVTZ-MOD      &  0.04865 &  0.10980 &  0.11941    &    0.18362   & 0.20720   & 0.21847        &  -0.01773  &  0.06700 &  0.08788  \\
  AVQZ-MOD      &  0.04822 &  0.11138 &  0.12140    &    0.18393   & 0.21025   & 0.22373        &  -0.01785  &  0.07163 &  0.09513  \\
  AV5Z-MOD      &  0.04806 &  0.11182 &  0.12113    &    0.18396   & 0.21169   & 0.22663        &  -0.01798  &  0.07351 &  0.09776  \\
Estimated CBS   &  0.04806 &  0.11245 &  0.12101    &    0.18396   & 0.21318   & 0.22967        &  -0.01798  &  0.07563 &  0.10068  \\
\hline                                                                                      
 {Other works}  & \multicolumn{3}{c|}{} & \multicolumn{3}{c|}{} & \multicolumn{3}{c|}{}  \\
 Estimated Exact$^b$ & \multicolumn{3}{c|}{0.12490} & \multicolumn{3}{c|}{} & \multicolumn{3}{c|}{}  \\
 {VMC/DMC$^c$}  & \multicolumn{3}{c|}{0.12526(35)/0.12659(56)} & \multicolumn{3}{c|}{0.20752(25)/0.22874(66)}     &   \multicolumn{3}{c|}{0.08278(35)/0.10533(66)}       \\ 
 {MBPT$^d$}     & \multicolumn{3}{c|}{           }  &   \multicolumn{3}{c|}{0.22778}            &   \multicolumn{3}{c|}{0.09988}         \\ 
 MRCI$^e$       & \multicolumn{3}{c|}{0.12473    }  &   \multicolumn{3}{c|}{0.22840}            &   \multicolumn{3}{c|}{0.10312}        \\ 
\end{tabular} 
\caption{Calculations for the F, F$^-$ and PsF systems in the aug-cc-pVXZ-mod (AVXZ-MOD) basis sets with ${\mathrm{X}=\text{T},\text{Q},\text{5}}$. For each basis set and system, we report the RHF total energies together with the correlation energies $\ecorr$ both at the CISD and LCCSD levels. 
Estimated CBS are also reported for correlation energies as extrapolations using Eq.\eqref{eq:cisd_cbs} with ${\mathrm{X}=\text{5}}$ , and we use the RHF values in the AV5Z-MOD basis sets for CBS mean-field energies. Electron affinities (EA), positron binding energy (PBE) and binding energy (BE) are also reported. \protect\linebreak
$^a$: a value of $-$0.25 au is taken as the internal energy of Ps. \protect\linebreak
$^b$: near exact values obtained from Ref. \onlinecite{exact_atoms}. \protect\linebreak
$^c$: results coming from the AGP-EPO trial wave functions of Ref. \onlinecite{Pos-QMC-ChaJorBarTka-JCTC-22}. \protect\linebreak
$^d$: MBPT results extrapolated at the CBS limit using the $\sum^{(2+\Gamma+3)}$ approximations for the self-energy of Ref. \onlinecite{Pos-GW-LudGri-10}. \protect\linebreak
$^e$: Multi-reference CI calculations extrapolated to the CBS limit of Ref. \onlinecite{Saito_2005}.
\label{table_PsF}}
\end{ruledtabular} 
\end{table}
\end{center}
\end{widetext}

The advantage of LCCSD over CISD is that it is free from size extensivity error, \textit{i.e.} it does not introduce 
an increasing error when increasing the number of correlated particles. 
Within our implementation, it consists in a series of CISD-like diagonalization, and the process is typically converged within five or six iterations. 
The choice of LCCSD is also motivated by the observation\cite{TauBar-JCP-09} that the LCCSD results, while neglecting exclusion of Pauli violation diagrams, 
are very competitive with their exact level of treatment, namely the coupled cluster with single and double substitution (CCSD).
The LCCSD is therefore known to perform accurately on molecular systems near ther equilibrium geometries, but discontinuities arise when bonds are stretched. 
Nevertheless, as the present calculations are performed atoms and their positronized versions, we do not expect to face such problems. 

In order to estimate and mitigate the incomplete basis set error, we employ the two-point CBS extrapolation as usually 
done in quantum chemistry\cite{HelKloKocNog-JCP-97,Halkier_1999}, which allows us to obtain reliable estimation of the 
correlation energy for each system near the CBS limit. The CISD correlation energy for a given basis set characterized by the cardinal number ${\mathrm{X}}$ is defined as 
\begin{equation}
\mathcal{E}_{\text{corr}}(\mathrm{X})=E_{\text{CISD}}(\mathrm{X})-E_{\text{RHF}}(\mathrm{X})<0, 
\end{equation}
and similarly for the LCCSD correlation energy, where one just replace 
$E_{\text{CISD}}(\mathrm{X})$ by $E_{\text{LCCSD}}(\mathrm{X})$. 

The CBS extrapolation method consists in assuming a power law for the correlation energy with respect to $\mathrm{X}$
\begin{equation}
\mathcal{E}_{\text{corr}}(\mathrm{X})=\mathcal{E}_{\text{corr}}^{\infty}+a\mathrm{X}^{-b}\label{Eq.onea},
\end{equation}
where $\mathcal{E}_{\text{corr}}^{\infty}$ is the unknown CBS limit of the correlation energy. 
By differentiating this equation at $\mathrm{X}$ and ${\mathrm{X}-1}$, one can eliminate the linear free parameter $a$ in Eq. (\ref{Eq.onea}) leading to 
\begin{equation}
\mathcal{E}_{\text{corr}}^{\infty}=\fracinlines{\mathcal{E}_{\text{corr}}(\mathrm{X})\mathrm{X}^b-\mathcal{E}_{\text{corr}}(\mathrm{X}\!-\!1)(\mathrm{X}\!-\!1)^b}{\mathrm{X}^b-(\mathrm{X}\!-\!1)^b}\label{Eq.oneb},
\end{equation}
which depends only on the non-linear free parameter $b$, which has been set to 3 based on benchmarks numerical tests\cite{HelKloKocNog-JCP-97,Halkier_1999}. One should notice that ${\mathcal{E}_{\text{corr}}^{\infty}}$ depends on the maximal cardinal number used for the extrapolation. We then estimate the CISD energy near the CBS limit using the following equation
\begin{equation}
\label{eq:cisd_cbs}
E_{\text{CISD}}^\infty\simeq E_\text{RHF}(\mathrm{X}\!=\!5)+\mathcal{E}_{\text{corr}}^{\infty},
\end{equation}
and similarly for the LCCSD energy. 

\begin{table}[htb]
\begin{ruledtabular}
\begin{tabular}{|l|@{\hskip -0.3in} c|@{\hskip -0.3in} c|@{\hskip -0.3in} c|}
Basis/System         &   F       & F$^-$     & PsF        \\                                                                          
\hline
AVTZ     & -99.40208 & -99.45081 &  -99.60883 \\
AVQZ     & -99.40921 & -99.45746 &  -99.62122 \\
AV5Z     & -99.41120 & -99.45926 &  -99.62768 \\
\end{tabular}   
\caption{RHF calculations for the F, F$^-$ and PsF systems in the usual augmented Dunning aug-cc-pVXZ (AVXZ) basis sets with ${\mathrm{X}=\text{T},\text{Q},\text{5}}$.\label{table-PsFAVXZ}}
\end{ruledtabular} 
\end{table}

In the following, we consider Eq. \eqref{eq:cisd_cbs} using the LCCSD correlation energy as the theoretical best estimate (TBE) when $\mathcal{E}_{\text{corr}}^{\infty}$ is evaluated from Eq. (\ref{Eq.oneb}) at ${\mathrm{X}=5}$. 

\section{Results\label{section-2}}

\subsection{Computational details\label{subsection-1}}
We applied the frozen-core approximations in all CISD and LCCSD calculations, which corresponds to a [He] and [Ne] core in the fluorine and chlorine based 
calculations, respectively. Regarding the AO basis sets considered here, we use modified versions of the standard aug-cc-pVXZ Dunning's family of basis sets with ${\mathrm{X}=\text{T},\text{Q},\text{5}}$, labelled here as ``aug-cc-pVXZ-mod" (AVX-MOD), with extra diffuse functions. The latter are constructed from the usual aug-cc-pVXZ basis sets by the following procedure: for each angular momentum in the AO basis, one adds three AO functions whose Gaussian exponents follow a geometric progression with one-half common ratio by taking, as reference, the smallest Gaussian exponent of the angular momentum considered. We implemented the extensions to positronic matter of the RHF, CISD and LCCSD approaches as a pluggin of the open-source, {\it Quantum Package} (QP) programming environment\cite{QP_2019}. The cross sections were calculated by interfacing the collision code with the QP software. Regarding now the energy differences involving the energy of the Ps atom, as the localised Gaussian basis functions used in this present work are inappropriate for this system, we simply consider the exact Ps binding energy of $-$0.25 au obtained by solving the hydrogenoid problem with a reduced mass ${\mu=\sfrac{1}{2}}$. 

\subsection{Benchmarking the \textit{Ab-initio} approach on PsF\label{subsection-2}}
Before studying the attachment of a positron to a chlorine anion, we begin our study by investigating the performance of our approaches on a smaller positronic halide, 
namely PsF, with other state-of-the-art methods to compute positronic systems. In the case of the PsF atom, our reference values are the diffusion Monte Carlo (DMC) results of Ref. \onlinecite{Pos-QMC-ChaJorBarTka-JCTC-22} obtained with refined correlation factors multiplied by an anti symmetrized geminal power together with electron-positron orbitals (AGP-EPO) trial wave functions 
fully optimized in Variational Monte Carlo (VMC). We also report the basis-set extrapolated MBPT\cite{Pos-GW-LudGri-10} and multi-reference CI\cite{Saito_2005} (MRCI) calculations for comparison with 
purely orbital-based \textit{ab initio} methods (\textit{i.e.} without any correlation factors). For a given chemical species A, we focus in the present work on the following energy differences: the electron affinity (EA) defined as EA=E(A) - E(A$^-$), the positron binding energy (PBE) defined as PBE=E(PsA) - E(A$^-$) and the binding energy (BE) defined as E(A) + E(Ps) - E(PsA). We report in Tab. \ref{table_PsF} the calculations of RHF total energies, CISD and LCCSD correlation energies together with the corresponding EA, PBE and BE in increasingly large AVXZ-MOD basis sets up to CBS extrapolation. We also compare with the MBT\cite{Pos-GW-LudGri-10} and MRCI\cite{Saito_2005} results 
together with the references QMC values\cite{Pos-QMC-ChaJorBarTka-JCTC-22} for these systems, which are also reported in Tab. \ref{table_PsF}. For comparison, we also performed RHF calculations in the standard aug-cc-pVXZ augmented Dunning basis sets (AVXZ) for these three species and present the values in Tab. \ref{table-PsFAVXZ}. 

Comparing first the RHF energy in the AVXZ and AVXZ-MOD basis sets, it clearly appears that the addition of extra diffuse functions in the latter basis sets allows for a smoother convergence of the total energy for all systems, including the positronic system. From the  quantitative point of view, while the difference in total energies between the AVQZ-MOD and AV5Z-MOD basis sets is of about 0.002 au for all species, it is considerably larger when using the standard Dunning basis sets only for the PsF system. The latter illustrates the need of additional diffuse functions to describe the density of the positron attached to the F$^-$ anion, which strongly suggests the use of the AVXZ-MOD basis sets instead of the standard augmented Dunning basis sets. 

Regarding now the behaviour of the correlation energy with the basis set, from Tab. \ref{table_PsF} it appears that it converges at a slower rate for the PsF system than for F and F$^-$. From a quantitative point of view, the difference in correlation energy between the AVQZ-MOD and AV5Z-MOD basis set is of about 0.006 au for F and F$^-$ while it is of about 0.009 au for PsF. The latter highlights the slower convergence of the electron-positron correlation effects with respect to the usual electron-electron correlation counterpart. 

Focussing now on the energy differences, from Tab. \ref{table_PsF} two global observations can be made: i) the correlation contribution to the EA, PBE and BE is remarkably large compared to the RHF contribution to these energy differences, ii) there is a significant difference between the CISD and LCCSD values. Comparing the EA at CISD and LCCSD levels with the near exact non relativistic values of Ref. \onlinecite{exact_atoms}, the error with the CBS extrapolated LCCSD values is less than 0.004 au while the error is multiplied by 3 with the CBS extrapolated CISD results. 

Focussing now on the PBE, we can notice that the CBS extrapolated CISD results have an error of about 0.016 au with respect to the DMC results, while CBS extrpolated LCCSD results agree with the DMC with less than 0.001 au, and similarly for the MBPT and MRCI results. Eventually one can notice that both LCCSD and MBPT slightly underestimates the BE of about 0.004 au with respect to DMC, while the underestimation is of about 0.002 au using the extrapolated MRCI. Overall, this study on the PsF system shows that the post-HF methods, such as our LCCSD approach and the MRCI of Sato\cite{Saito_2005}, 
together with the MBPT methods, once extrapolated near the CBS limit, manage to accurately describe the electron-positron correlation effects playing a differential role in the BE and PBE even if they do not use any correlation factor. 

\subsection{\textit{Ab-initio} study of the stability of PsCl\label{subsection-3}}
\begin{figure}[tp!]
\centering
\includegraphics[scale=1.05]{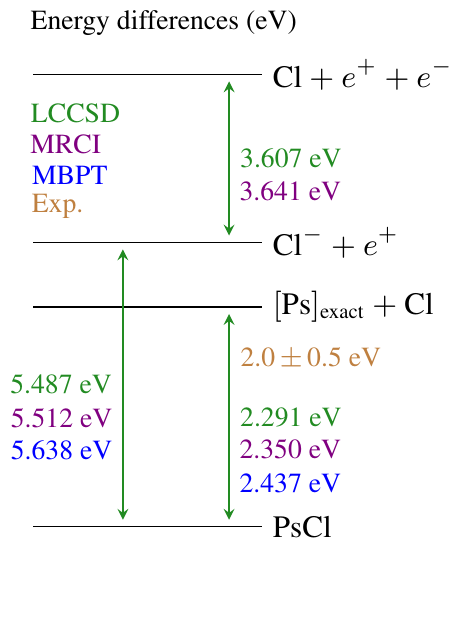}
\vskip-8mm
\caption{Energy level diagrams of PsCl computed with: (green) our TBE estimates of the LCCSD; (violet) the MRCI reference values of Saito {\it et al.}\cite{Saito_2005}; (blue) the MPBT values of Gribakin {\it et al.}\cite{Pos-GW-LudGri-10}; (brown) experimental value of the Ps binding energy (BE) in PsCl\cite{Tao_1969}. For the dissociative states, only the thresholds (zero relative kinetic energy) are shown. \label{fig_delta_e}} 
\end{figure}

We continue our study by investigating the energy differences associated with the species involved in the various dissociation channels of PsCl, in a similar way that what was done in subsection \ref{subsection-2} for PsF. We report in Fig. \ref{fig_delta_e} and Tab. \ref{table_PsCl} the results obtained with our \textit{ab initio} approaches, together with the existing MRCI\cite{Saito_2005} and MBPT\cite{Pos-GW-LudGri-10}. In Fig. 2, the EA, PBE and BE energy differences are depicted in unit of electronvolt. In Tab. \ref{table_PsCl} the results draw similar conclusions than on the study of PsF: there is a slower convergence of the correlation energy with respect to the basis set for the positronic system, and there is a significant difference between the CISD and LCCSD results. Comparing now with the preexisting values in the literature for the energy differences,  
we notice that our CBS extrapolated LCCSD results are in reasonable agreement with the MRCI results (within 0.002 au for the EA, 0.001 au for the PBE, and 0.0035 au for the BE) and the MBPT results 
(within 0.0065 au for the PBE and within 0.005 au for the BE). We can then conclude that our CBS extrapolated LCCSD results are reliable for the various energy differences computed on the PsCl system. 

\begin{widetext}
\begin{center}
\begin{table}[htb]
\begin{ruledtabular}
\begin{tabular}{|l|c c c| c c c| c c c|}
  Basis/System        &\multicolumn{3}{c|}{Cl}             &\multicolumn{3}{c|}{Cl$^-$}                & \multicolumn{3}{c|}{PsCl}            \\ 
\hline
                &RHF       & $\ecorr$(CISD) &  $\ecorr$(LCCSD)  &   RHF       &  $\ecorr$(CISD) &  $\ecorr$(LCCSD)     &  RHF       & $\ecorr$(CISD) &  $\ecorr$(LCCSD) \\ 
\hline
 AVTZ-MOD       &-459.48027 &-0.18289       &-0.19712           &-459.57349   &-0.21153         &-0.23092              &-459.71650  &-0.23923        &-0.27317  \\
 AVQZ-MOD       &-459.48331 &-0.19617       &-0.21180           &-459.57635   &-0.22802         &-0.24948              &-459.71944  &-0.25916        &-0.29910  \\
 AV5Z-MOD       &-459.48380 &-0.20086       &-0.21699           &-459.57679   &-0.23356         &-0.25558              &-459.71988  &-0.26660        &-0.30956  \\
Estimated CBS   &-459.48380 &-0.20578       &-0.22243           &-459.57679   &-0.23937         &-0.26198              &-459.71988  &-0.27441        &-0.32053  \\
\hline
Energy difference &\multicolumn{3}{c|}{EA = E(Cl)-E(Cl$^-$)} &    \multicolumn{3}{c|}{PBE = E(Cl$^-$)-E(ClPs)}   &\multicolumn{3}{c|}{BE = E(Cl) + E(Ps)$^a$ - E(ClPs)} \\
\hline
                &  RHF     &  CISD    &  LCCSD      &    RHF       & CISD      & LCCSD          &   RHF      &  CISD    &  LCCSD    \\
  AVTZ-MOD      &0.09322   &0.12186   &0.12702      &0.14301       &0.17071    &0.18526         &-0.01377    &0.04257   &0.06228    \\
  AVQZ-MOD      &0.09304   &0.12489   &0.13072      &0.14309       &0.17423    &0.19271         &-0.01387    &0.04912   &0.07343    \\
  AV5Z-MOD      &0.09299   &0.12569   &0.13158      &0.14309       &0.17613    &0.19707         &-0.01392    &0.05182   &0.07865    \\
 Estimated CBS  &0.09299   &0.12658   &0.13254      &0.14309       &0.17813    &0.20164         &-0.01392    &0.05471   &0.08418    \\
\hline                                                                                      
 {Other works}  & \multicolumn{3}{c|}{} & \multicolumn{3}{c|}{} & \multicolumn{3}{c|}{}  \\
 Estimated Exact$^b$ & \multicolumn{3}{c|}{0.133} & \multicolumn{3}{c|}{} & \multicolumn{3}{c|}{}  \\
 {MBPT$^c$}     & \multicolumn{3}{c|}{           }  &   \multicolumn{3}{c|}{0.20718}            &   \multicolumn{3}{c|}{0.08956}         \\ 
 MRCI$^d$       & \multicolumn{3}{c|}{0.13381    }  &   \multicolumn{3}{c|}{0.20256}            &   \multicolumn{3}{c|}{0.08636}        \\ 
\end{tabular}   
\caption{Calculations for the Cl, Cl$^-$ and PsCl systems in the aug-cc-pVXZ-mod (AVXZ-MOD) basis sets with ${\mathrm{X}=\text{T},\text{Q},\text{5}}$. For each basis set and system, we report the RHF total energies together with the correlation energies $\ecorr$ both at the CISD and LCCSD levels. 
Estimated CBS are also reported for correlation energies as extrapolations using Eq.\eqref{eq:cisd_cbs} with $\mathrm{X}=5$ and $\mathrm{X}=\text{Q}$, and we use the RHF values in the AV5Z-MOD 
basis sets for CBS mean-field energies.  ²
Electron affinities (EA), positron binding energy (PBE) and binding energy (BE) are also reported. \protect\linebreak
$^a$: a value of -0.25 au is taken as the internal energy of Ps. \protect\linebreak
$^b$: near exact values obtained from Ref. \onlinecite{exact_atoms}. \protect\linebreak
$^c$: MBPT results extrapolated at the CBS limit using the $\sum^{(2+\Gamma+3)}$ approximations for the self-energy of Ref. \onlinecite{Pos-GW-LudGri-10}. \protect\linebreak
$^d$: Multi-reference CI calculations extrapolated to the CBS limit of Ref. \onlinecite{Saito_2005}.\label{table_PsCl}}
\end{ruledtabular} 
\end{table}
\end{center}
\end{widetext}

\begin{figure}[bp!]
\centering
\begin{subfigure}[t]{0.449\textwidth}
\hskip-1mm\includegraphics[width=1\textwidth]{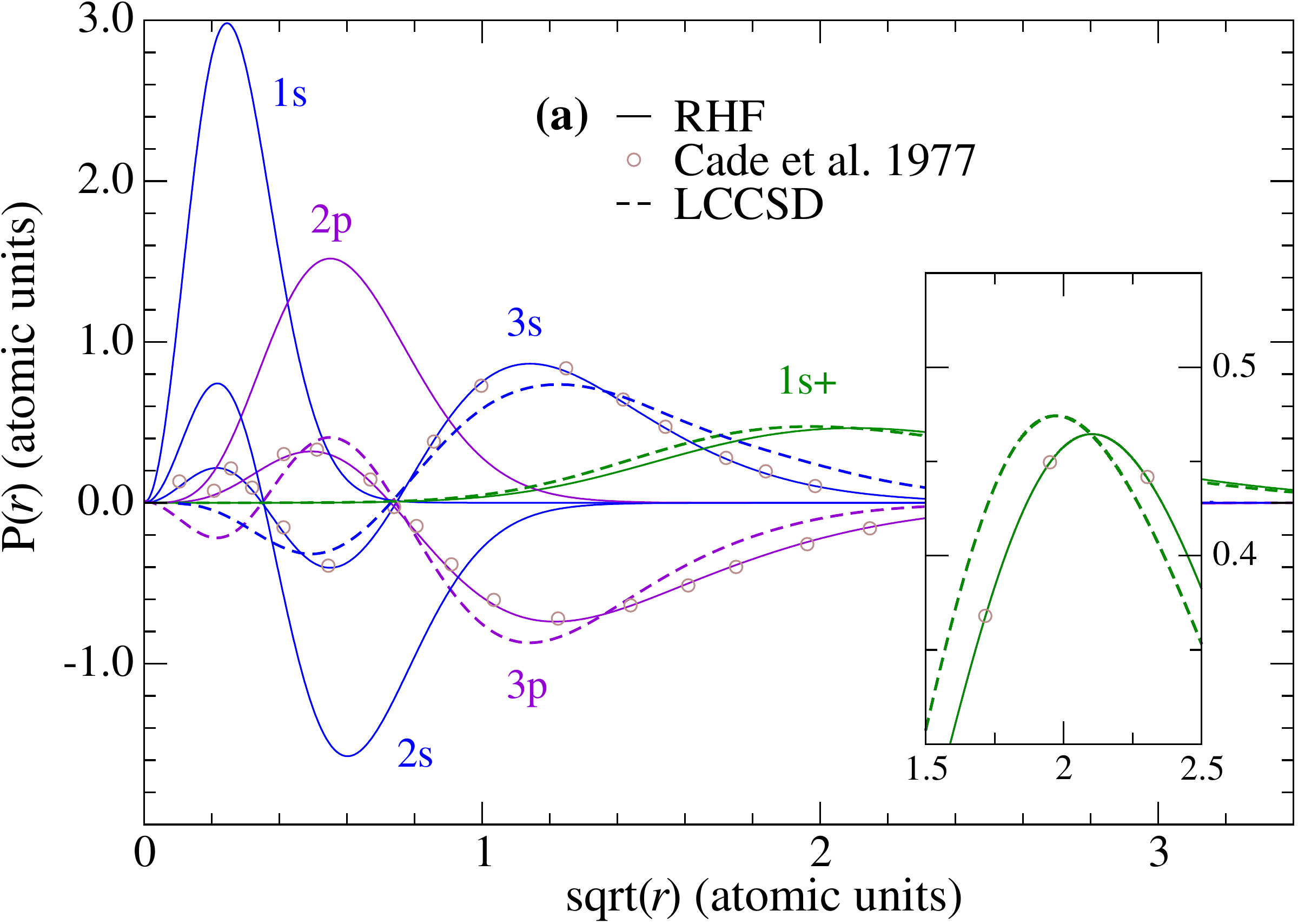}
\caption{\label{data-1}}
\end{subfigure}
\vskip-6mm
\begin{subfigure}[t]{0.45\textwidth}
\includegraphics[width=1\textwidth]{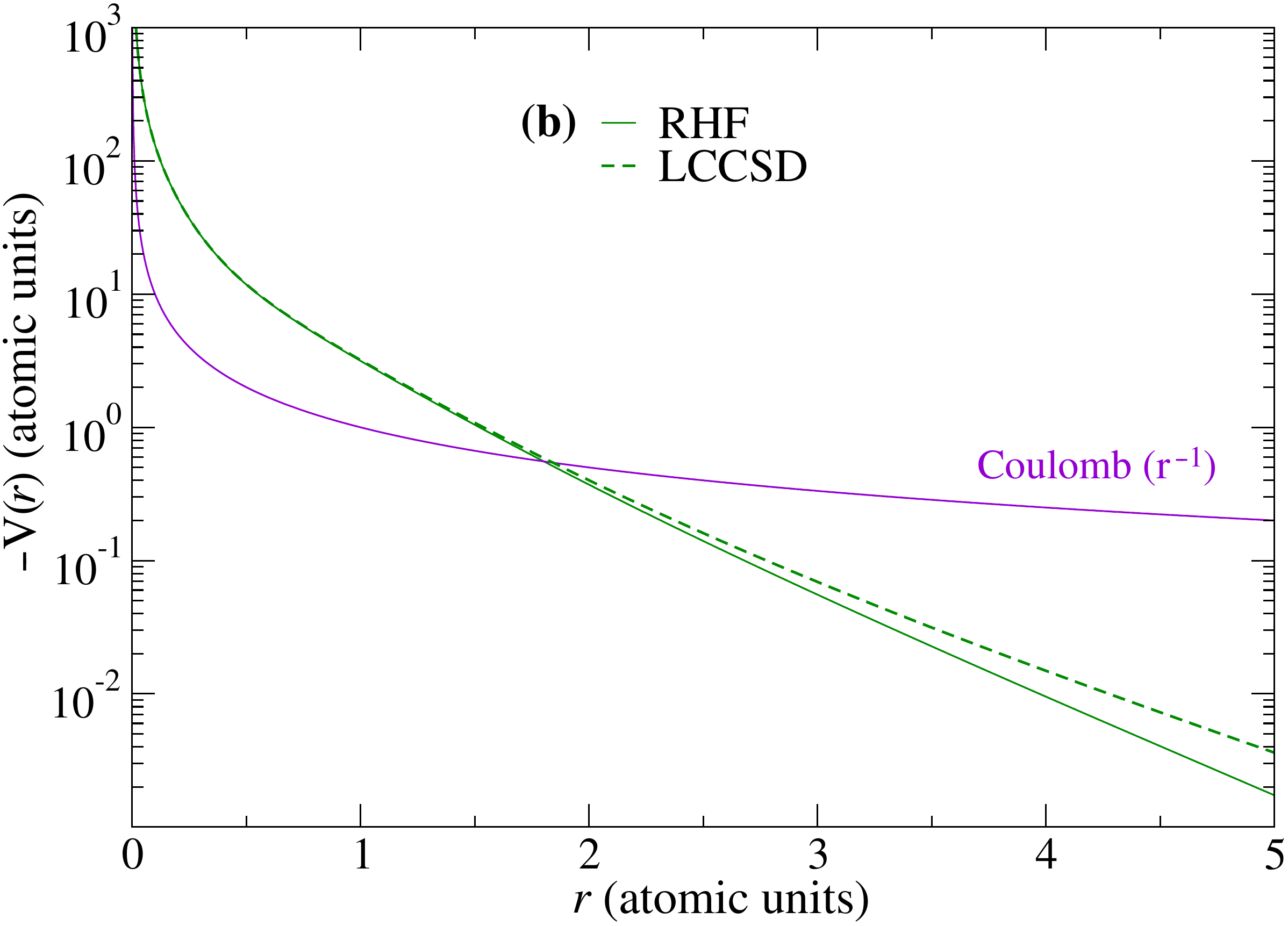}
\caption{\label{data-2}}
\end{subfigure}
\vskip-7mm
\caption{Structure of PsCl at the AV5Z-MOD level of description. \textbf{(a)} Evolution of the SPPs of the radial wavefunctions $P_{n\ell}(r)=rR_{n\ell}(r)$ of the electrons (for the $1s\!-\!3p$ orbitals) and of the positron (for the lowest energy orbital) bound in PsCl as functions of the distance $r$ from nucleus. The following approaches are employed: (continuous line) RHF; (dashed line) LCCSD. (Circle) SPPs of the HF radial wavefunctions of the electrons and of the positron bound in PsCl extracted from Ref. \onlinecite{Cade_1977}. \textbf{(b)} Evolution of the absolute value of the square brackets term in Eq. (\ref{es.1}) as a function of the distance from the nucleus.\label{data}} 
\end{figure}

\subsection{PsCl production cross sections\label{subsection-4}}

\begin{figure}[tp!]
\centering
\begin{subfigure}[t]{0.45\textwidth}
\includegraphics[width=1\textwidth]{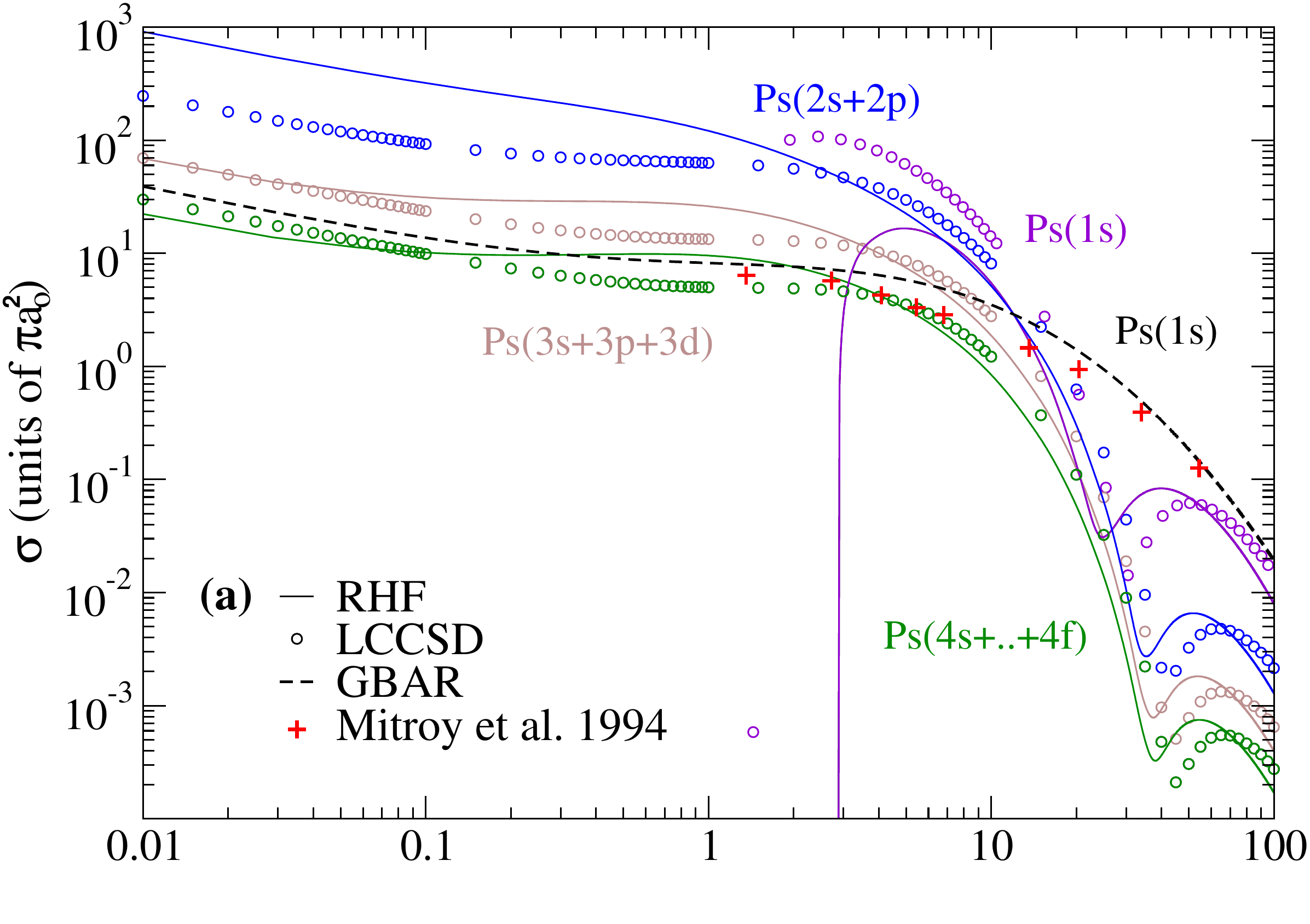}
\caption{\label{CBA}}
\end{subfigure}
\vskip-8mm
\begin{subfigure}[t]{0.45\textwidth}
\includegraphics[width=1\textwidth]{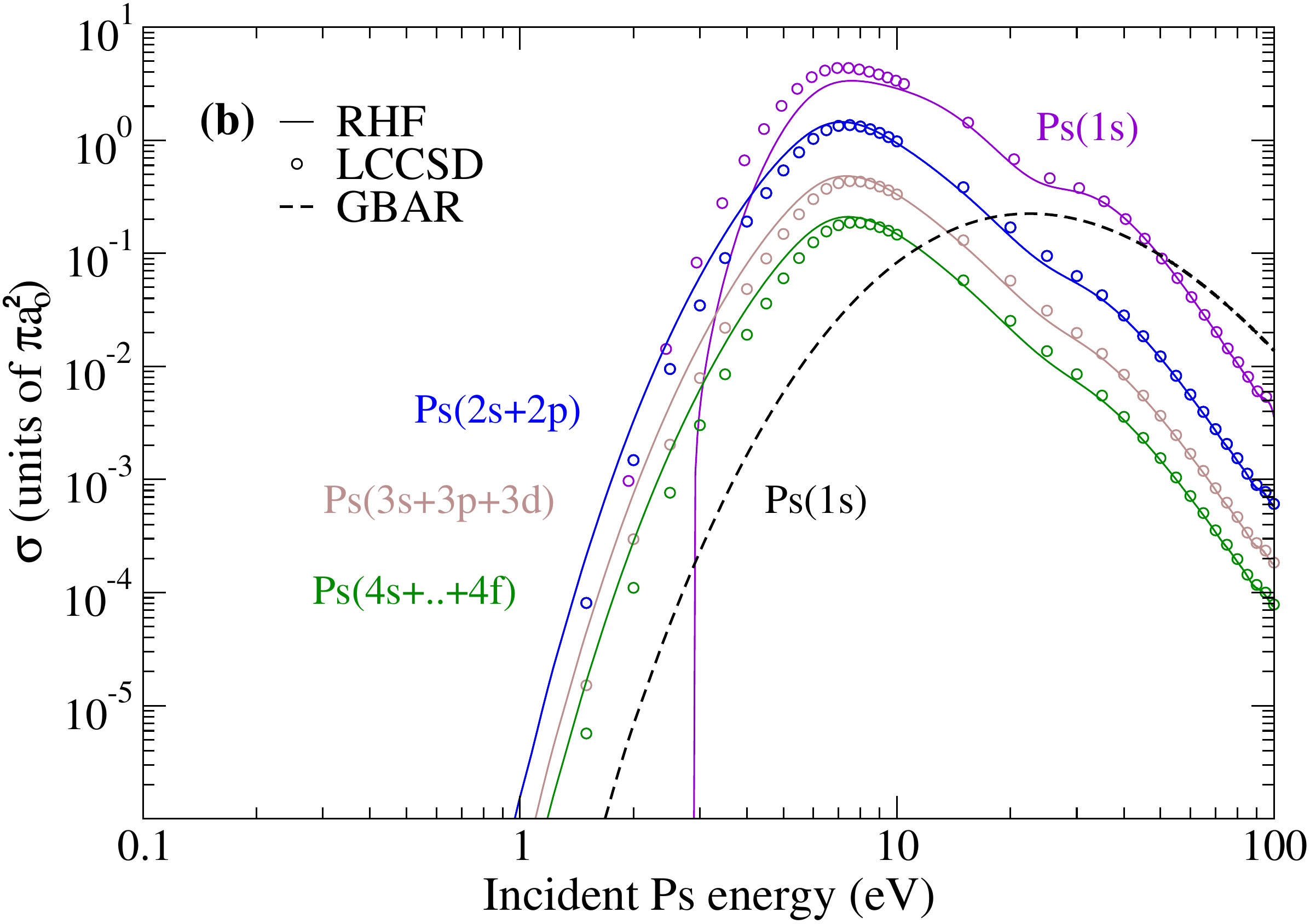}
\caption{\label{CDW}}
\end{subfigure}
\vskip-6mm
\caption{Evolution of the partial cross sections associated with the main channels (${n\leqslant 4}$) of reaction (\ref{eqn:the_reaction}), summed over $\ell$ as functions of the Ps impact energy using the AV5Z-MOD basis set. In the formulation: \textbf{(a)} CBA; \textbf{(b)} CDW. The data associated with PsCl involving these two basis sets are generated in the following approximations: (continuous line) RHF; (circle) LCCSD. Hydrogen-like radial wavefunctions for Ps and 26 partial waves to ensure the convergence of the transition amplitude are considered. In the framework of GBAR experiment: (dashed line) $\overline{\mathrm{H}}(1s)$ production cross sections from the ${\text{Ps}(1s)+\overline{p}}$ charge exchange transfer in CBA/CDW formulation; (plus) Unitarized Born Approximation (UBA)\cite{Mitroy_1994} cross sections for the same reaction.\label{CS}}
\end{figure}

The computation of the PsCl production cross section relies on the \textit{ab-initio} results obtained above with the many-body electron-positron wavefunctions. The two frames of Fig.
\ref{data} depict quantities derived from the PsCl wavefunction involved in the cross section calculations. These data were generated using the AV5Z-MOD basis set in both RHF and LCCSD approximations. The SPPs of the radial wavefunctions of the frozen/active electrons and of the positron bound in PsCl are shown in Fig. \ref{data}\subref{data-1}. At the level of RHF approximation, we find that these SPPs of the radial wavefunctions are in good agreement with those obtained by Cade \textit{et al}.\cite{Cade_1977}. The slight difference between the SPP of the LCCSD radial wavefunction of the positron bound to Cl$^-$ and the one obtained in the RHF approximation reflects a weakly correlated system. For core electrons, the SPPs of the radial wavefunctions of electrons bound in PsCl are identical in both approximations, as expected in the frozen core approximation. The perturbative potential for the charge exchange transfer in PsCl production is reported in Fig. \ref{data}\subref{data-2}, and we verify numerically that the potential i) is asymptotically null and ii) tends faster towards zero than the Coulomb interaction which takes place between an antiproton and an electron in the case of GBAR.

The partial cross sections for Ps states for ${n\leq 4}$ are displayed in Figs. \ref{CS}\subref{CBA}-\subref{CDW} for CBA and CDW models. These cross sections are given in units of ${\pi a_0^2=\text{0.88}\!\times\!\text{10}^{-\text{16}}}$ cm$^2$ as a function of the Ps impact energy. For both models considered here, we observe that cross sections  decrease as a function of Ps impact energy. This illustrates the fact that the capture probability of the positron by the negatively charged target decreases as the Ps impact energy increases. In addition, both models predict that the cross sections decrease when considering final excited states of Ps. Except for the  Ps(${n=1}$) channel where the reaction threshold lies at 2.88 eV (RHF) and 1.44 eV (LCCSD), respectively, we emphasize that all channels are open at low Ps impact energies. In this energy range, the behaviour of CBA cross sections is different from those of CDW cross sections. Indeed, the inclusion of Coulomb distortions in the entrance channel leads to a collapse of the cross sections at low Ps impact energy (Fig. \ref{CS}\subref{CDW}). 

By analogy with GBAR, we thereafter turn our attention to the CBA cross sections, as the comparison between experimental data\cite{Merrison_1997} and previous results obtained by Comini \textit{et al.}\cite{Comini_Merrisson_2014} in the case of ground state positronium suggest that CBA is more reliable than CDW. 

Within the 3$-$10 eV impact energy range, the CBA results (Fig. \ref{CS}\subref{CBA}) show that the consideration of a chloride anion target leads to larger cross sections than an antiproton target (GBAR) when the positronium is in its ground state (${n=1}$). Since the incident velocity of the Ps distribution expected in future experiments will be of the order of $10^5$ m.s$^{-1}$ (see Ref. \onlinecite{Antonello_2020}), which corresponds to kinetic energies below 100 meV, the incoming channel Ps(${n=2}$) which dominates at low Ps impact energy is of particular interest. This shows first that laser excitation of Ps is experimentally required to achieve the production of PsCl, as the  Ps(${n=1}$) channel is closed in this low energy range. In second instance, we specify that due to the experimental constraints the $2s$ metastable state of Ps is often favoured over the $2p$ state. This is because the ${2s}$ state of Ps, which can be produced by driving the ${1^3s\!-\!3^3p}$ and ${3^3p\!-\!2^3s}$ transitions\cite{Antonello_2019}, has a long annihilation lifetime. In the configuration ortho (o-Ps), the latter is in fact eight times larger than those of the ground state\cite{Cooke_2015} (${8\!\times\!142}$ ns). 

In terms of numerical convergence, mismatches are observed between the cross sections involving the RHF and LCCSD data associated with PsCl. These discrepancies illustrate the importance of correlation in the structure and reactivity of the positronium chloride system. For instance, the CBA partial cross section for Ps(${n=3}$) computed from the RHF data is equal to 26.04 ${\pi a_0^2}$ (resp. ${\text{1.75}\!\times\!\text{10}^{-\text{3}}}$ ${\pi a_0^2}$) for an impact energy of 1 eV (resp. 50 eV), while when LCCSD data are used we obtain 13.33 ${\pi a_0^2}$ (resp. ${\text{0.78}\!\times\!\text{10}^{-\text{3}}}$ ${\pi a_0^2}$), respectively. The differences between RHF and LCCSD  values for the CBA cross sections can be attributed to the change in i) the positron binding energy or ii) the shape of the SPP of the radial wavefunction of the positron bound to Cl$^-$. In order to understand which effect is dominant, we recalculated the CBA cross sections by combining on the one hand the RHF binding energy and the SPP of the LCCSD radial wavefunction of the bound positron, and on another hand the LCCSD binding energy and the SPP of the RHF radial wavefunction. This approach is justified when considering the AV5Z-MOD basis set which induces a negligible error on the effective short-range perturbative potential (see Fig. \ref{data}\subref{data-2}). The results are shown in App. \ref{annexe-1} for Ps(${n=2}$). It establishes that the cross sections discrepancies arise from the difference on positron binding energies between RHF and LCCSD for Ps impact energies below 10 eV, and from the difference on the SPPs of radial wavefunctions of the positron for a higher impact energies. This is due to the fact that at low impact energies, the Ps kinetic energy ${v^2}$ involved in the energy conservation law 
\begin{equation}
\fracinlines{k_{\beta}^2}{2}=v^2+\epsilon_n-\epsilon_{1s\scalemath{0.8}{+}},
\end{equation}
takes values on the same order of magnitude than the positron binding energy ${\epsilon_{1s\scalemath{0.8}{+}}}$, as reported in Tab. \ref{table_PsCl}, noting that below 10 eV, ${v^2<\text{0.37}}$ au. It results that for a given Ps binding energy ${\epsilon_n=-1/(4n^2)}$, the ejected electron energy is strongly modified by a small variation of the positron binding energy. However, this is no longer true at larger impact energies, for which the changes on the SPP of the positron radial wavefunction, negligible at low Ps impact energies, contribute to the differences of the cross sections.

Without the inclusion of Coulomb distortions (Fig. \ref{CS}\subref{CBA}), we also note that a minimum appears in each partial cross section for Ps impact energy in the range of 25$-$42 eV, depending on the internal state of Ps. Since for a given Ps binding energy and basis set the RHF and LCCSD approximations lead to a shift of the order of 5 eV of these minima, we conclude that the latter are not driven by the energy conservation law, but come from the discrepancies on the SPP of the positron radial wavefunction. Moreover, the absence of minimum in the GBAR cross section computed for Ps($1s$) suggests a Cooper minimum-like behaviour\cite{Cooper_1962} of the CBA partial cross sections, resulting from the electronic structure of the Cl$^-$ target considered here. 

In Figs. \ref{CS-par}\subref{1}-\subref{4} we represent the convergence of the CBA partial cross sections as a function of the AVXZ-MOD basis sets. According to the arguments mentioned above, these results show a faster convergence of the cross sections computed with the RHF data than those computed with the LCCSD data. Indeed, if we consider as a reference in Tab. \ref{table_PsCl} the positron binding energies obtained with the AV5Z-MOD basis set, we find that the one obtained with the TZ basis set accounts for 99.94$\%$ of the RHF value, against only 94.01$\%$ of the LCCSD value. In order to obtain an estimate of the cross sections that can be provided beyond the LCCSD approximation, we finally substituted in the most accurate available data associated with PsCl ({\it i.e.} the modified basis set of cardinal number ${\mathrm{X}=5}$) the positron binding energy obtained near the CBS limit. This corresponds to the dashed lines shown in Figs. \ref{CS-par}. As this TBE value of the positron binding energy of 5.487 eV is comparable to the one obtained by Saito \textit{et al}.\cite{Saito_2003} (5.512 eV), and that the discrepancies in the cross sections respectively computed with the AV5Z-MOD and TBE positron binding energies remain small, we can deduce that the correlation effects in PsCl are accurately taken into account for the cross section calculations.

\begin{figure}[bp!]
\centering
\hskip-5mm
\begin{subfigure}[bp!]{0.46\textwidth}
\includegraphics[width=1.05\textwidth]{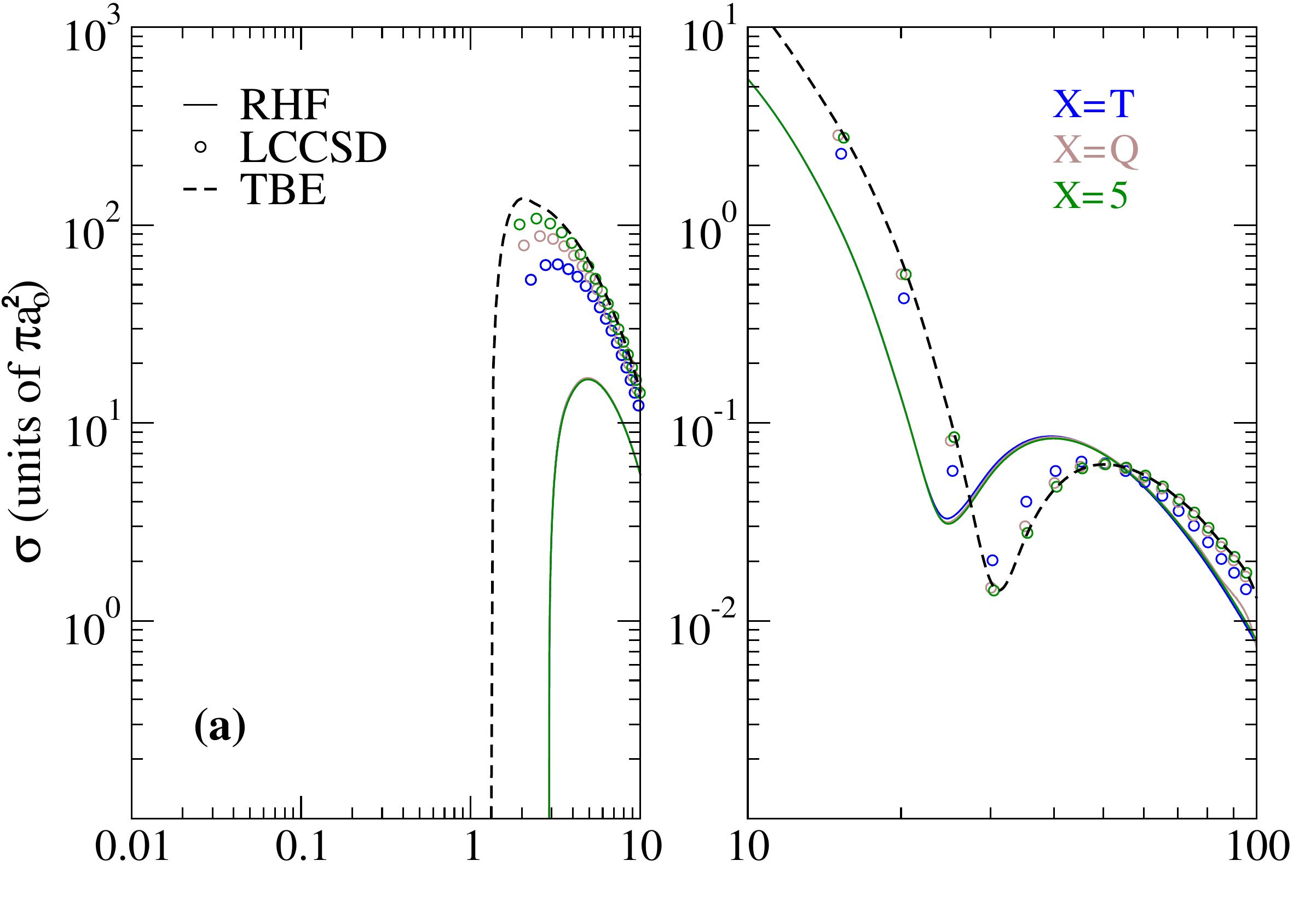}
\end{subfigure}
\vskip-3mm
\hskip-5mm
\begin{subfigure}[t]{0.46\textwidth}
\includegraphics[width=1.05\textwidth]{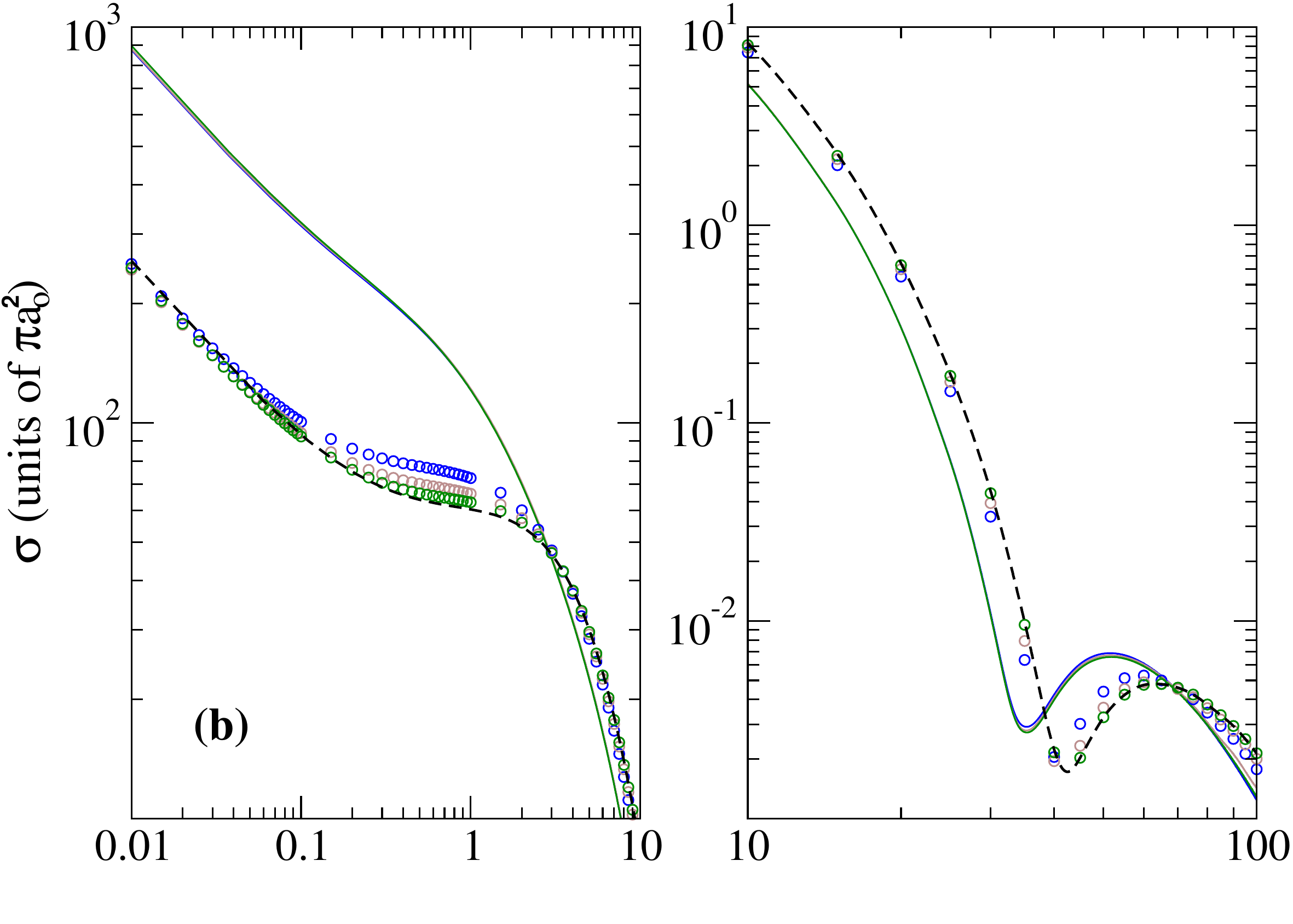}
\end{subfigure}
\vskip-3mm
\hskip-5mm
\begin{subfigure}[tp!]{0.46\textwidth}
\includegraphics[width=1.05\textwidth]{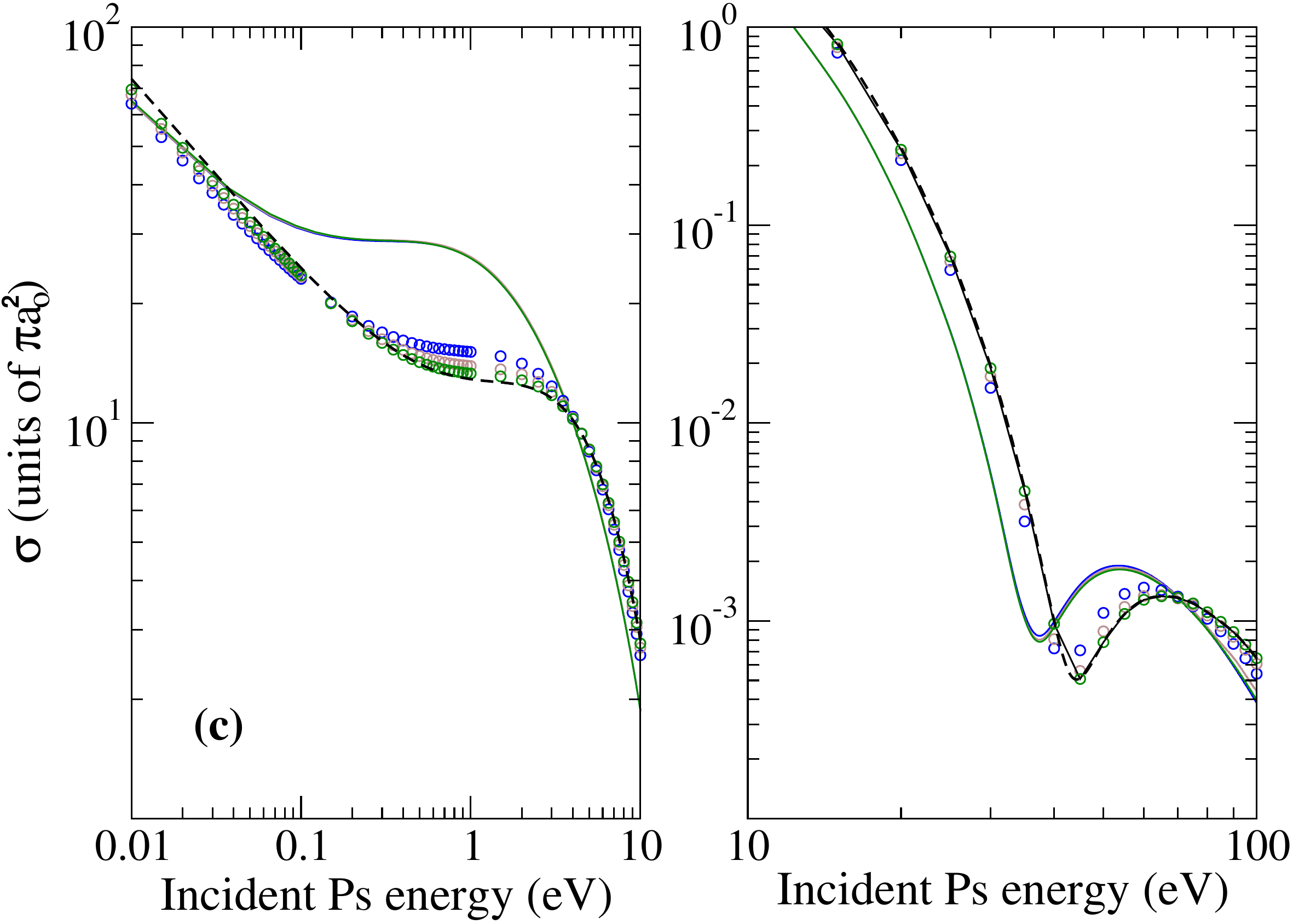}
\end{subfigure}
\end{figure}

\setcounter{figure}{5}

\begin{figure}[tp!]
\centering
\hskip-5mm
\begin{subfigure}[t]{0.46\textwidth}
\includegraphics[width=1.05\textwidth]{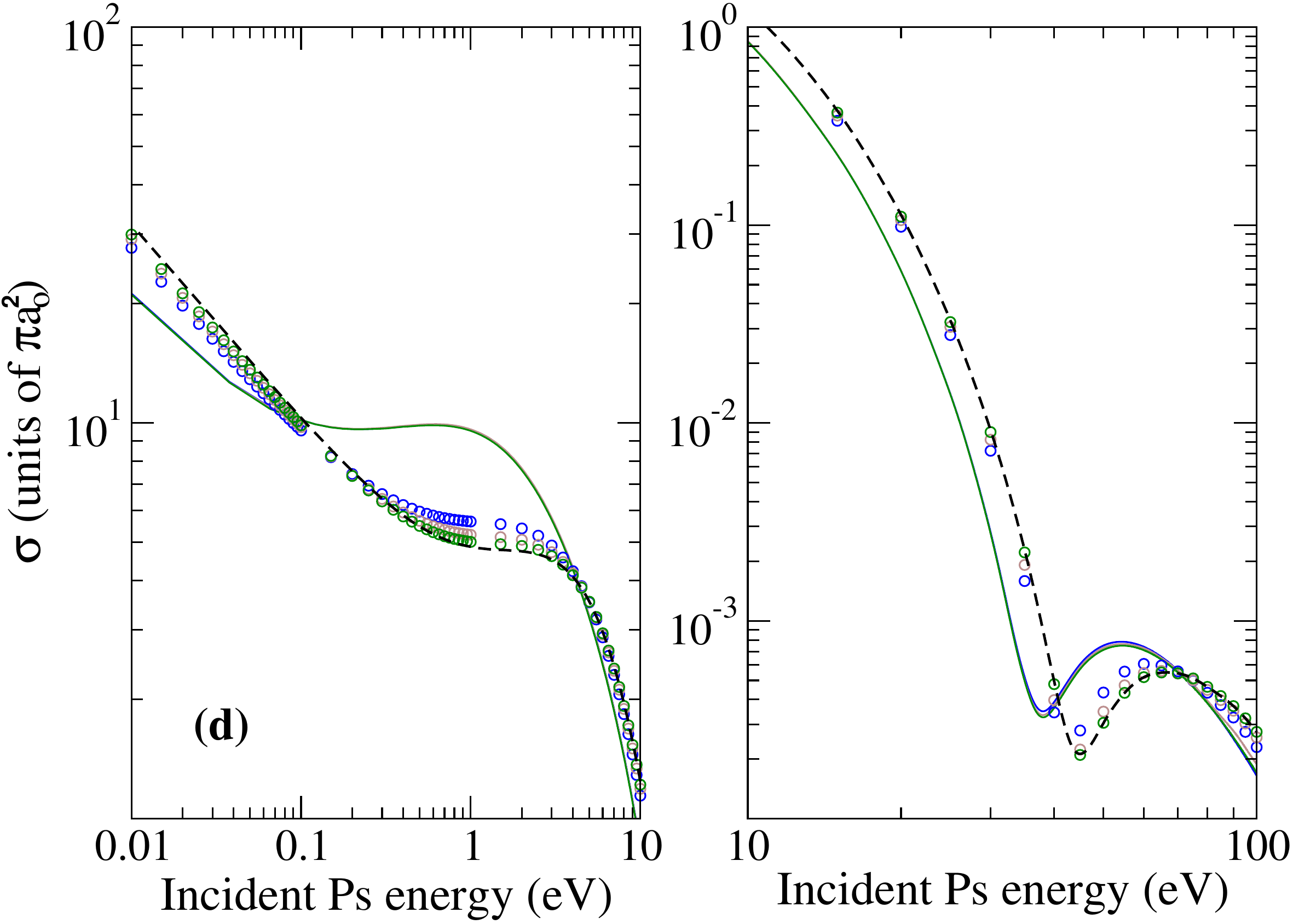}
\vskip-6mm
\caption{\label{1}}
\caption{\label{2}}
\caption{\label{3}}
\caption{\label{4}}
\end{subfigure}
\vskip-19mm
\caption{Evolution of the CBA partial cross sections as a function of Ps impact energy for: \textbf{(a)} ${n=1}$; \textbf{(b)} ${n=2}$; \textbf{(c)} ${n=3}$; \textbf{(d)} ${n=4}$. The data associated with PsCl are computed with the AVXZ-MOD basis sets for ${\mathrm{X}=\text{T},\text{Q},\text{5}}$ in the following approximations: (continuous lines) RHF; (circle) LCCSD; (dashed line) TBE cross sections involving LCCSD approximation and the most accurate positron binding energy obtained near the CBS limit ({\it i.e.} $-$0.20164 au). For readability purpose, we used different incident Ps energy scales in the 0.01$-$10 eV (left frames) and in the 10$-$100 eV (right frames) ranges.\label{CS-par}}
\end{figure}

\section{Conclusions and perspectives\label{section-3}}

In the present work, we performed a theoretical study on the possibility to produce positronium chloride in its ground state from charge exchange between Ps and Cl$^-$. For this purpose, perturbative approaches based on the FBA in its three-body formulation have been employed, using an effective short-range perturbative potential that allows dealing with the electronic structure of the chlorine anion target. The chemical structure of PsCl was modelled using \textit{ab-initio} methods at both RHF and LCCSD levels of approximations, adapted here to account for the presence of a positron in the electronic cloud. 
We benchmarked the accuracy of the present approaches on a smaller positronic halide, namely PsF, by comparing with state-of-the-art QMC calculations, together with MRCI and MBPT. We found that, once extrapolated to the CBS limit, our LCCSD energy differences on PsF agree with the QMC references within a few 0.001 au. Regarding the PsCl system, the LCCSD energy differences extrapolated near the CBS limit are comparable to those of other methods such as MBPT and MRCI. In addition, our approach allowed for a systematic study of the effect of electronic correlation on the partial cross sections. The latter have been generated with a collision code relying on \textit{ab-initio} positronic wavefunctions, electronic densities and energy differences as input. The results have shown that cross sections are significantly affected by the electronic correlation in PsCl.

As a perspective concerning the structure calculations on PsCl, the annihilation rate could be evaluated by first benchmarking the electron-positron contact density on positronic systems already studied, using the extrapolation technique proposed by Ludlow \textit{et al.} This will allow to compare the accuracy of the lifetimes derived from our approach with those of QMC methods including a correlation factor in the wavefunction. Regarding now the collisional aspect, it would be interesting to deal with the projectile-target interaction at the second order of the Born series\cite{Briggs_1980}. Indeed, a more accurate estimation of cross sections, especially at low Ps impact energies, will be obtained. At a later stage, charge exchanges between Ps and molecules such as Cl$_2$ can also be investigated. This can be achieved numerically as the QP software allows to deal with molecules, in constrast to the use of the B-spline basis functions, and that the orientation of the molecules can be taken into account in the effective short-range perturbative potential\cite{Caillat_2006}. Experimentally, this direction would make it possible to determine which of the Cl$^-$ or Cl$_2$ targets is optimal for the production of positronium chloride.

\section*{Acknowledgements}
This research work was supported by an Emergence Sorbonne University funding.
\normalem

%
\appendix

\section{Computation of PsCl production cross sections in CDW-IS approximation\label{appendix-1}}

In this part, details of analytical cross sections calculations using CDW-IS (Initial State) approximation are presented. These developments are mostly taken from the calculations of the Refs. \onlinecite{Swann_2016} and \onlinecite{Comini_2013}, where the coordinates used for the reaction are those of Fig. \ref{Figure-1}. 

The \textit{post} form of the $S$-matrix elements, which corresponds to the central quantity related to the transition amplitude, are expressed as\cite{McDowell_1971} 
\begin{equation}
S_{\beta\alpha}=-i\!\int_{\mathbb{R}}\!dt\hspace{0.2mm}\bra{\Phi_{\beta}(t)}V_{\beta}\ket{\Phi_{\alpha}(t)},\label{es.7}
\end{equation}
where ${\ket{\Phi_{\alpha}(t)}}$ and ${\ket{\Phi_{\beta}(t)}}$ are the asymptotic states associated to the entrance and exit channel, respectively. In Eq. (\ref{es.7}), the expression of the effective short-range perturbative potential ${V_{\beta}}$ modelling the projectile-target interaction is given by the Eq. (\ref{es.1}). 

On the one hand, the wavefunction corresponding to the entrance channel is defined by
\begin{equation}
\hskip-2mm\Phi_{\alpha}(\boldsymbol{r},\boldsymbol{s},t)=e^{\scalemath{0.9}{-i\hspace{0.1mm}\epsilon_{n}t-v^2t}}\mathcal{F}_{\boldsymbol{k}_-}^{_{(+)}}(\boldsymbol{r})\mathcal{F}_{\boldsymbol{k}_+}^{_{(+)}}(\boldsymbol{s})\psi_{n\ell m}(\boldsymbol{q}),\label{es.4}
\end{equation}
where ${\epsilon_n}$ and ${v^2}$ are respectively the binding and kinetic energies of the incident Ps. The wavefunctions of the electron bound in Ps atom, prepared in a quantum state ${(n\ell m})$, is defined by ${\psi_{n\ell m}(\boldsymbol{q})=R_{n\ell}(q)Y_{\ell m}(\hat{\boldsymbol{q}})}$. The Coulomb wavefunctions ${\mathcal{F}_{\boldsymbol{k}_-}^{_{(+)}}(\boldsymbol{r})}$ and ${\mathcal{F}_{\boldsymbol{k}_+}^{_{(+)}}(\boldsymbol{r})}$ of Eq. (\ref{es.4}) describe the propagation of the incoming electron and positron in the continuum of the target. 

i) In the case of GBAR (antiproton target), their complete expressions are analytical and given by 
\begin{align}
&\mathcal{F}_{\boldsymbol{k}_-}^{_{(+)}}(\boldsymbol{r})=e^{\scalemath{0.9}{-\alpha_-\!\scalemath{1.2}{\sfrac{\pi}{2}}}}\hspace{0.4mm}\Gamma(1+i\alpha_-)\nonumber\\
&\phantom{\mathcal{F}_{\boldsymbol{k}_-}^{(+)}(\boldsymbol{r})=}\times\hspace{0.2mm}_1\hspace{-0.4mm}F_1(-i\alpha_-;1; i(k_- r-\boldsymbol{k}_-\hspace{-0.3mm}\cdot\hspace{-0.3mm}\boldsymbol{r}))e^{\scalemath{0.9}{\hspace{0.2mm}i\boldsymbol{k}_-\hspace{-0.3mm}\cdot\hspace{-0.3mm}\boldsymbol{r}}},\label{es.11}\\[1mm]
&\mathcal{F}_{\boldsymbol{k}_+}^{_{(+)}}(\boldsymbol{s})=e^{\scalemath{0.9}{-\alpha_+\!\scalemath{1.2}{\sfrac{\pi}{2}}}}\hspace{0.4mm}\Gamma(1+i\alpha_+)\nonumber\\
&\phantom{\mathcal{F}_{\boldsymbol{k}_+}^{(+)}(\boldsymbol{s})=}\times\hspace{0.2mm}_1\hspace{-0.4mm}F_1(-i\alpha_+;1; i(k_+ s-\boldsymbol{k}_+\hspace{-0.3mm}\cdot\hspace{-0.3mm}\boldsymbol{s}))e^{\scalemath{0.9}{\hspace{0.2mm}i\boldsymbol{k}_+\hspace{-0.3mm}\cdot\hspace{-0.3mm}\boldsymbol{s}}}\label{es.12},
\end{align}
where ${\boldsymbol{k}_{\mp}\simeq\sfrac{1}{2}\hspace{0.2mm}\boldsymbol{k}_{\alpha}}$. In the above equations, the incident Ps wavevector is defined by ${\boldsymbol{k}_{\alpha}=2v\hat{\boldsymbol{\text{z}}}}$. The Sommerfeld parameters ${\alpha_{\mp}=\mp Z_T/v}$ associated to Eqs. (\ref{es.11}) and (\ref{es.12}) are defined as a function of a net charge of the antiproton target ${Z_T=-1}$. It should be noted that if no distortions are included in the entrance channel (${\alpha_{\mp}=0}$), the Coulomb Born approximation (CBA) is obtained from the present CDW-IS approximation. 

ii) In the present case (chlorine anion target), the so-called shooting method\cite{Goldberg_1981} is employed to numerically determine the Coulomb wavefunctions of Eq. (\ref{es.4}). The first step of this method consists in the resolution of the radial Schr\"odinger equations describing a negative(positive) charged particle moving in the effective potential induced by Cl$^-$. For this purpose, the fourth-order Runge-Kutta integration scheme is employed. In the framework of the local-density approximation\cite{Hervieux_2019}, the radial Schr\"odinger equations needed to be solved are expressed as
\begin{align}
&\bigg[\fracinline{d^2}{dr^2}+2\Big(E_{\text{kin}}-V_{\text{eff}}^{(\ell)}(r)\Big)\bigg]P_{k_{\mp},\ell}(r)=0,\label{es.20}\\[1.5mm]
&V_{\text{eff}}^{(\ell)}(r)=\fracinline{\ell(\ell+1)}{2r^2}+V(r),\\[1.5mm]
&V(\boldsymbol{r})=\pm\bigg[\int_{\mathbb{R}^3}\!\!\!d\boldsymbol{r}'\fracinlines{\rho_{\alpha}(\boldsymbol{r}')}{|\boldsymbol{r}-\boldsymbol{r}'|}-\fracinlines{Z_A}{|\boldsymbol{r}|}\bigg]+V_{\text{XC}}[\rho_{\alpha}(\boldsymbol{r})]\label{es.19},\\[1.5mm]
&V_{\text{XC}}[\rho_{\alpha}(\boldsymbol{r})]=-\Big(\fracinliness{3}{\pi}\rho_{\alpha}(\boldsymbol{r})\Big)^{\!\sfrac{1}{3}}\nonumber\\
&\phantom{V_{\text{XC}}[\rho_{\alpha}(\boldsymbol{r})]=}-0.0333\ln\bigg[1+11.4\Big(\fracinline{4\pi}{3}\rho_{\alpha}(\boldsymbol{r})\Big)^{\!\sfrac{1}{3}}\bigg],\label{es.21}
\end{align}
where in Eq. (\ref{es.20}) ${P_{k_{\mp},\ell}(r)/r}$ are the radial parts of the Coulomb wavefunctions of degree $\ell$, and ${E_{\text{kin}}=k_{\mp}^2/2}$ is the kinetic energy of the electron(positron). In Eq. (\ref{es.19}), both signs occur according to the negative(positive) charge of the incoming electron(positron). The exchange-correlation (XC) potential (Eq. (\ref{es.21})), which is directly parametrized from the one-electron density ${\rho_{\alpha}}$ of the chlorine anion\cite{Gunnarsson_1976}, is expressed as the sum of two terms: i) the first one is exactly derivable by a variational approach from the Hartree-Fock exchange energy and ii) the second one corresponding to the correlation potential is not driven by the HF formalism. In a second step, the Str\"omgren method\cite{Seaton_1962,Caillat_2015} is employed to normalize the radial parts of the Coulomb wavefunctions. As these normalizations are made on the energy scale, the solutions obtained from the integration scheme are multiplied by ${\sqrt{\pi}(2E_{\text{kin}})^{-\sfrac{1}{4}}}$, in order to remain consistent with the normalization adopted in the collision code through the COULFG \cite{COULFG_1982} subroutine. To illustrate the present shooting method on the PsCl production cross sections, the deviations of the CDW cross sections obtained in subsection \ref{subsection-4} from those involving the potentials used in GBAR (${V(\boldsymbol{r})\simeq\mp Z_T/|\boldsymbol{r}|}$ for $r\rightarrow\infty$) are shown in Appendix \ref{annexe-2} for the ground state of Ps.

On the other hand, the wavefunction corresponding to the exit channel is defined by
\begin{align}
\hskip-2mm&\phantom{\Phi_{\beta}(\boldsymbol{r},\boldsymbol{s},t)=e}^{\scalemath{0.9}{-i\hspace{0.1mm}\epsilon_{1s\scalemath{0.8}{+}}t-\sfrac{i}{2}\hspace{0.2mm}k_{\beta}^2\hspace{0.2mm}t+i\hspace{0.2mm}\boldsymbol{k}_{\beta}\hspace{-0.3mm}\cdot\hspace{-0.3mm}\boldsymbol{r}}}\psi_{1s\scalemath{0.8}{+}}(\boldsymbol{s})\nonumber,\\[-5.5mm]
\hskip-2mm&\Phi_{\beta}(\boldsymbol{r},\boldsymbol{s},t)=e\label{es.5}
\end{align}
where ${\boldsymbol{k}_{\beta}}$ is the wavevector of the ejected electron. The single-particle picture of the ground state wavefunction of the positron bound in PsCl and the associated binding energy are respectively defined by ${\psi_{1s\scalemath{0.8}{+}}(\boldsymbol{s})=R_{1s\scalemath{0.8}{+}}(s)Y_{10}(\hat{\boldsymbol{s}})}$ and ${\epsilon_{1s\scalemath{0.8}{+}}}$. According to the neutral charge of the produced PsCl (who takes on the role of $\overline{\mathrm{H}}$ in GBAR), the propagation of the ejected electron is described by a plane wave. 

Using Eq. (\ref{es.4}) and Eq. (\ref{es.5}) allows to recast the $S$-matrix elements (\ref{es.7}) in the form
\begin{align}
&S_{\beta\alpha}=-2i\pi\delta\big(\sfrac{1}{2}\hspace{0.2mm}k_{\beta}^2++\epsilon_{1s\scalemath{0.8}{+}}\!-v^2\!-\epsilon_{n}\big)T_{\beta\alpha},\label{es.6}\\[1mm]
&T_{\beta\alpha}=\int_{\mathbb{R}^{^6}}\!\!d\boldsymbol{r}d\boldsymbol{s}\hspace{0.2mm}e^{\scalemath{0.9}{-i\boldsymbol{k}_{\beta}\hspace{-0.3mm}\cdot\hspace{-0.3mm}\boldsymbol{r}}}\complexconjugate{\psi}_{1s\scalemath{0.8}{+}}(\boldsymbol{s})V_{\beta}(\boldsymbol{r},\boldsymbol{s})\nonumber\\[1mm]
&\phantom{T_{\beta\alpha}=}\times\mathcal{F}_{\boldsymbol{k}_-}^{(+)}(\boldsymbol{r})\mathcal{F}_{\boldsymbol{k}_+}^{(+)}(\boldsymbol{s})\psi_{n\ell m}(\boldsymbol{r}-\boldsymbol{s})\label{es.18},
\end{align}
where ${T_{\beta\alpha}}$ is the transition amplitude. The energy law conservation of the collision process is ensured by the $\delta$-function appearing in the Eq. (\ref{es.6}). The latter comes from the integration over time in the $S$-matrix elements. At first order of the Born series, the differential cross section for a given magnetic quantum numbers of the Ps atom can be derived from the Fermi's golden rule\cite{Swann_2016}
\begin{equation}
d\sigma_{n\ell m}=\fracinlines{2\pi}{j}|T_{\beta\alpha}|^2\delta\big(\sfrac{1}{2}\hspace{0.2mm}k_{\beta}^2+\epsilon_{1s\scalemath{0.8}{+}}\!-v^2\!-\epsilon_{n}\big)d\rho_{\beta}.\label{es.9}
\end{equation}
In this equation, ${j=\sfrac{1}{2}\hspace{0.2mm}k_{\alpha}}$ is the flux density of incident Ps and ${d\rho_{\beta}=(2\pi)^{-3}d\boldsymbol{k}_{\beta}}$ is the density of final states. Using the spherical coordinates ${(k_{\beta},\theta,\varphi)}$ for the representation of the ${\boldsymbol{k}_{\beta}}$ vector in phase space, one has 
\begin{equation}
d\boldsymbol{k}_{\beta}\hspace{-0.5mm}=k_{\beta}d(k_{\beta}^2/2)d\boldsymbol{\hat{k}}_{\beta},
\end{equation}
with the solid angle ${d\boldsymbol{\hat{k}}_{\beta}\hspace{-0.5mm}=\sin\theta\hspace{0.1mm}d\theta\hspace{0.1mm}d\varphi}$. Finally, the total cross section is obtained by averaging over the magnetic quantum numbers of Ps and by performing the integration over ${d\boldsymbol{k}_{\beta}}$ in Eq. (\ref{es.9}). This leads to
\begin{equation}
\sigma_{n\ell}=\fracinline{k_{\beta}}{k_{\alpha}}(2\pi^2\hat{\ell})^{-1}\!\!\sum_{m=-\ell}^\ell\!|T_{\beta\alpha}|^2,\label{es.8}
\end{equation}
where the notation ${\hat{\ell}\equiv2\ell+1}$ is employed. It is worth mentioning that since both Ps and PsCl atoms are composed of a single positron, there is no need to include spin degree of freedom in Eq. (\ref{es.8}). Furthermore, the following multipole expansion 
\begin{align}
&\mathcal{F}_{\boldsymbol{k}_-}^{_{(+)}}(\boldsymbol{r})=4\pi\sum_{\ell_im_i}i^{\ell_i}e^{\scalemath{0.9}{\hspace{0.2mm}i\delta_{\ell_i}}}\scalemath{0.98}{\bigg[}\fracinliness{P_{k_-,\ell_i}(r)}{r}\scalemath{0.98}{\bigg]}\nonumber\\[1mm]
&\phantom{\mathcal{F}_{\boldsymbol{k}_-}^{_{(+)}}(\boldsymbol{r})=}\times\complexconjugate{Y}_{\ell_im_i}(\hat{\boldsymbol{k}}_-)Y_{\ell_im_i}(\hat{\boldsymbol{r}}),\label{es.16}
\end{align}
of the Coulomb wavefunction for the electron is employed to numerically compute the transition amplitude. In Eq. (\ref{es.16}), the overall phase shift including Coulombic and non-Coulombic effects is defined by ${\delta_{\ell_i}}$. The phase components resulting form the non-Coulombic effects are determined using the Wronskian method (see eg Ref. \onlinecite{Caillat_2015}). Similarly, the multipole expansion of the Coulomb wavefunction for the positron is obtained by solving Eq. (\ref{es.20}) with the minus sign in front of the bracket term of Eq. (\ref{es.19}), and by substituting in Eq. (\ref{es.16}) the set of variables ${(k_-,\ell_i,m_i,\boldsymbol{r})}$ by the set of variables ${(k_+,\ell_i',m_i',\boldsymbol{s})}$. 

As the integration variables (${\boldsymbol{r},\boldsymbol{s}}$) appearing in ${T_{\beta\alpha}}$ are those of the laboratory frame, a separation of variables into the wavefunctions $\psi_{n\ell m}(\boldsymbol{r}-\boldsymbol{s})$ must be performed. For the angular part, we use the following multipole expansion\cite{Franz_1992}
\begin{align}
\hskip-2.5mm&Y_{\ell m}(\boldsymbol{\hat{q}})=\scalemath{1.2}{\Big[}\fracinliness{(-)^m}{q}\scalemath{1.2}{\Big]}^{\ell}\sum_{\lambda\mu}(-)^{\lambda}\sqrt{\hat{\ell}!\hat{\ell}4\pi}\scalemath{1.25}{\big[}\hat{\lambda}!\hspace{0.4mm}\big(\widehat{\ell-\lambda}\big)!\scalemath{1.25}{\big]}^{-\sfrac{1}{2}}r^{\ell-\lambda}\nonumber\\[1mm]
\hskip-2.5mm&\phantom{Y_{\ell m}(\boldsymbol{\hat{q}})=}\times s^{\lambda}\scalemath{0.82}{
\begin{pmatrix}
\hspace{0.2mm}\ell-\lambda& \!\lambda &\!\ell\hspace{-1.6mm}\phantom{a}\\[1mm]
\hspace{0.2mm}m-\mu&\!\mu&\!-m\hspace{-1.6mm}\phantom{a}
\end{pmatrix}}Y_{\ell-\lambda m-\mu}(\boldsymbol{\hat{r}})\hspace{0.4mm}Y_{\lambda \mu}(\boldsymbol{\hat{s}}),\label{es.17}
\end{align}
where ${0\leqslant\lambda\leqslant \ell}$ and ${|\mu|\leqslant\lambda}$. Concerning the radial part, the projection of the scalar product into the complete basis set of Legendre polynomial is used, leading to  
\begin{align}
&V_{\beta}(\boldsymbol{r},\boldsymbol{s})q^{-l}R_{n\ell}(q)=4\pi\sum_{\ell_dm_d}\mathcal{J}_{n\ell;\ell_d}(r,s)\nonumber\\[1mm]
&\phantom{V_{\beta}(\boldsymbol{r},\boldsymbol{s})q^{-l}R_{n\ell}(q)=}\times\complexconjugate{Y}_{\ell_dm_d}(\hat{\boldsymbol{r}})Y_{\ell_dm_d}(\hat{\boldsymbol{s}}),\\
&\mathcal{J}_{n\ell;\ell_d}(r,s)=\sfrac{1}{2}\!\int\limits_{-1}^1\!d\xi V_{\beta}(\boldsymbol{r},\boldsymbol{s})q^{-\ell}R_{n\ell}(q)P_{\ell_d}(\xi)\label{es.14}.
\end{align}
In Eq. (\ref{es.14}), the Legendre polynomials of degree ${\ell_d}$ are defined by ${P_{\ell_d}}$ and the modulus of the $q$-vector is defined by ${q=\surd(r^2+s^2-2\hspace{0.1mm}r\hspace{0.1mm}s\hspace{0.1mm}\xi)}$. In addition, the ${q^{-\ell}}$ factor appearing in the integral comes from the Eq. (\ref{es.17}). It should be noted that the above relations are derived from each other by using the addition theorem and the completeness relation for spherical harmonics. With the help of Eqs. (\ref{es.16}), (\ref{es.17}) and (\ref{es.14}), Eq. (\ref{es.18}) can be recast in the from
\begin{align}
&T_{\beta\alpha}=(4\pi)^{\sfrac{3}{2}}(-)^{\ell}(\hat{\ell}!\hspace{0.2mm}\hat{\ell})^{\sfrac{1}{2}}\hspace{-1.2mm}\sum_{\ell_f\ell_d\ell_i\ell_i'}\hspace{-1.2mm}i^{\hspace{0.2mm}\ell_i+\ell_i'-\ell_f}e^{\scalemath{0.9}{i(\delta_{\ell_i}+\delta_{\ell_i'})}}\nonumber\\[-1mm]
&\phantom{T_{\beta\alpha}=}\times\sum_{\lambda=0}^{l}\mathcal{A}_{\ell m;\lambda \ell_d\ell_i\ell_i'\ell_f}\mathcal{R}_{n\ell;\lambda \ell_d\ell_i\ell_i'\ell_f},\label{es.13}
\end{align}
where the expression of the angular/radial coefficients in eq. (\ref{es.13}) are given by
\begin{widetext}
\begin{align}
&\mathcal{A}_{\ell m;\lambda \ell_d\ell_i\ell_i'\ell_f}=
\fracinline{(-)^{\hspace{0.2mm}\lambda}\hat{\ell}_d\hspace{0.4mm}\hat{\ell}_i\hspace{0.4mm}\hat{\ell}_i'\hspace{0.4mm}\hat{\ell}_f^{\hspace{0.4mm}\sfrac{1}{2}}}{\big((2\lambda)!(2(\ell-\lambda))!\big)^{\hspace{-0.2mm}\sfrac{1}{2}}}
\sum_{\chi}\hat{\chi}\hspace{0.2mm}
\scalemath{0.9}{
\begin{pmatrix}
\ell_f&\ell-\lambda&\chi\\[1mm]
0&0&0
\end{pmatrix}
\begin{pmatrix}
\ell_i&\ell_d&\chi\\[1mm]
0&0&0
\end{pmatrix}
\begin{pmatrix}
\ell_i'&\ell_d&\lambda\\[1mm]
0&0&0
\end{pmatrix}}
\nonumber\\[1mm]
&\phantom{\mathcal{A}_{\ell m;\lambda \ell_d\ell_i\ell_i'\ell_f}=}
\times\scalemath{0.9}{
\sum_{\mu=-\lambda}^{\lambda}
\begin{pmatrix}
\ell-\lambda&\lambda&\ell\\[1mm]
m-\mu&\mu&-m
\end{pmatrix}
\begin{pmatrix}
\ell_f&\ell-\lambda&\chi\\[1mm]
-m&m-\mu&\mu
\end{pmatrix}
\begin{pmatrix}
\ell_i&\ell_d&\chi\\[1mm]
0&\mu&-\mu
\end{pmatrix}
\begin{pmatrix}
\ell_i'&\ell_d&\lambda\\[1mm]
0&-\mu&\mu
\end{pmatrix}
},\\[2mm]
&\mathcal{R}_{n\ell;\lambda \ell_d\ell_i\ell_i'\ell_f}=\int_{\mathbb{R}^2_+}\!\!\!\hspace{-0.2mm}drds\hspace{0.2mm}r^{\hspace{0.2mm}\ell-\lambda+1}s^{\lambda+1}P_{k_-\ell_i}(r)j_{\ell_f}(k_{\beta}r)\mathcal{J}_{n\ell;\ell_d}(r,s)P_{k_+\ell_i'}(s)R_{1s\scalemath{0.8}{+}}(s).\label{es.15}
\end{align}
For our simulations, the integrals (\ref{es.15}) were computed using the method of Gauss-Laguerre quadrature, with a number of nodes equal to 96. To ensure the convergence of the transition amplitude, the upper limits of the partial waves $\ell_i$, $\ell_i'$ and $\ell_f$ are set to 26, and the other upper limits are set by the selection rules derived from the Wigner coefficients.

\setcounter{figure}{0}
\renewcommand{\thefigure}{\Alph{section}}

\section{Additional figures\label{appendix-2}}

\vskip-2mm
\onecolumngrid
\begin{figure}[htb]
\centering
\begin{subfigure}[htb]{0.45\textwidth}
\includegraphics[width=1\textwidth]{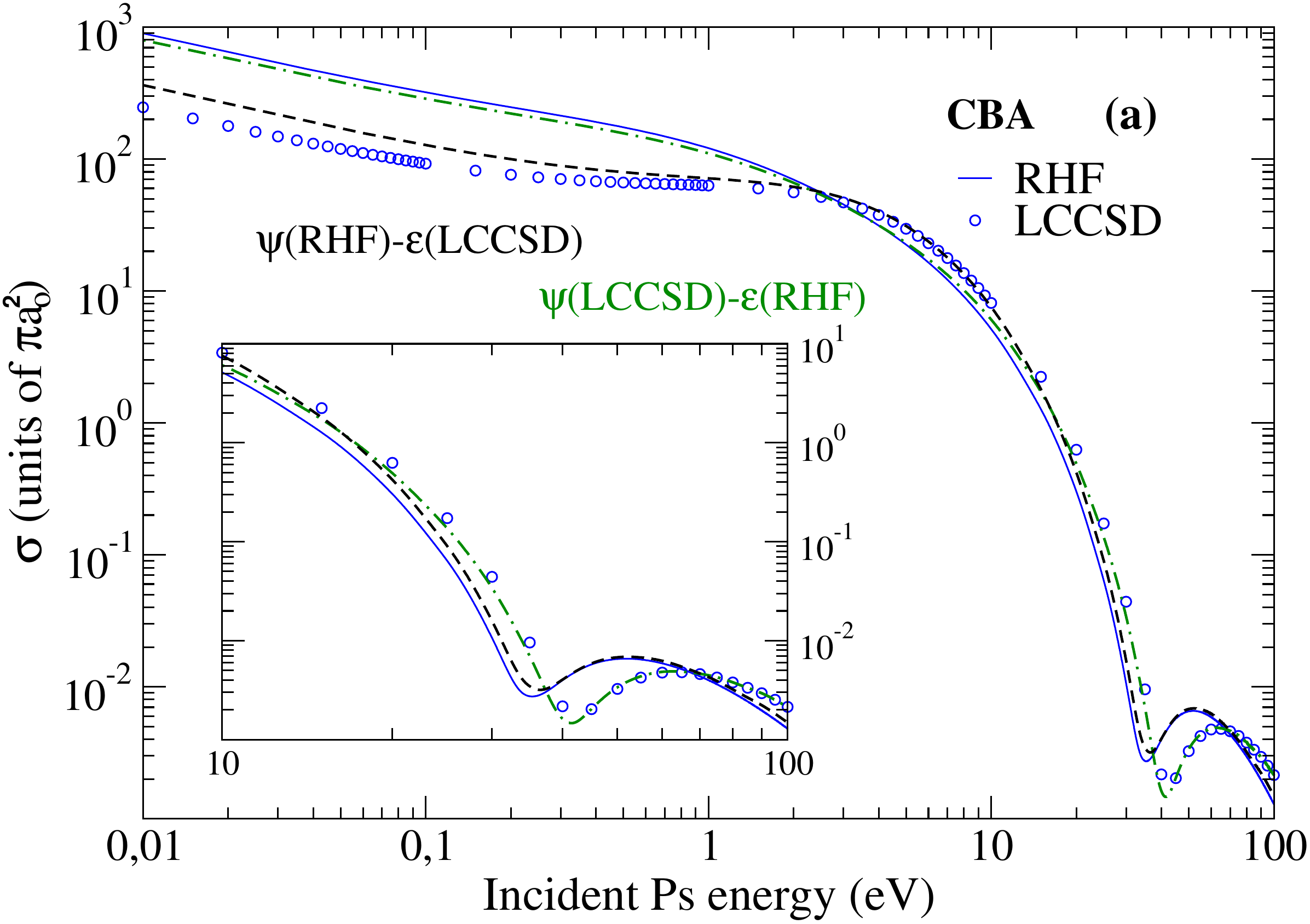}
\subcaption{\label{annexe-1}}
\end{subfigure}
\hskip+6mm
\begin{subfigure}[htb]{0.45\textwidth}
\includegraphics[width=1\textwidth]{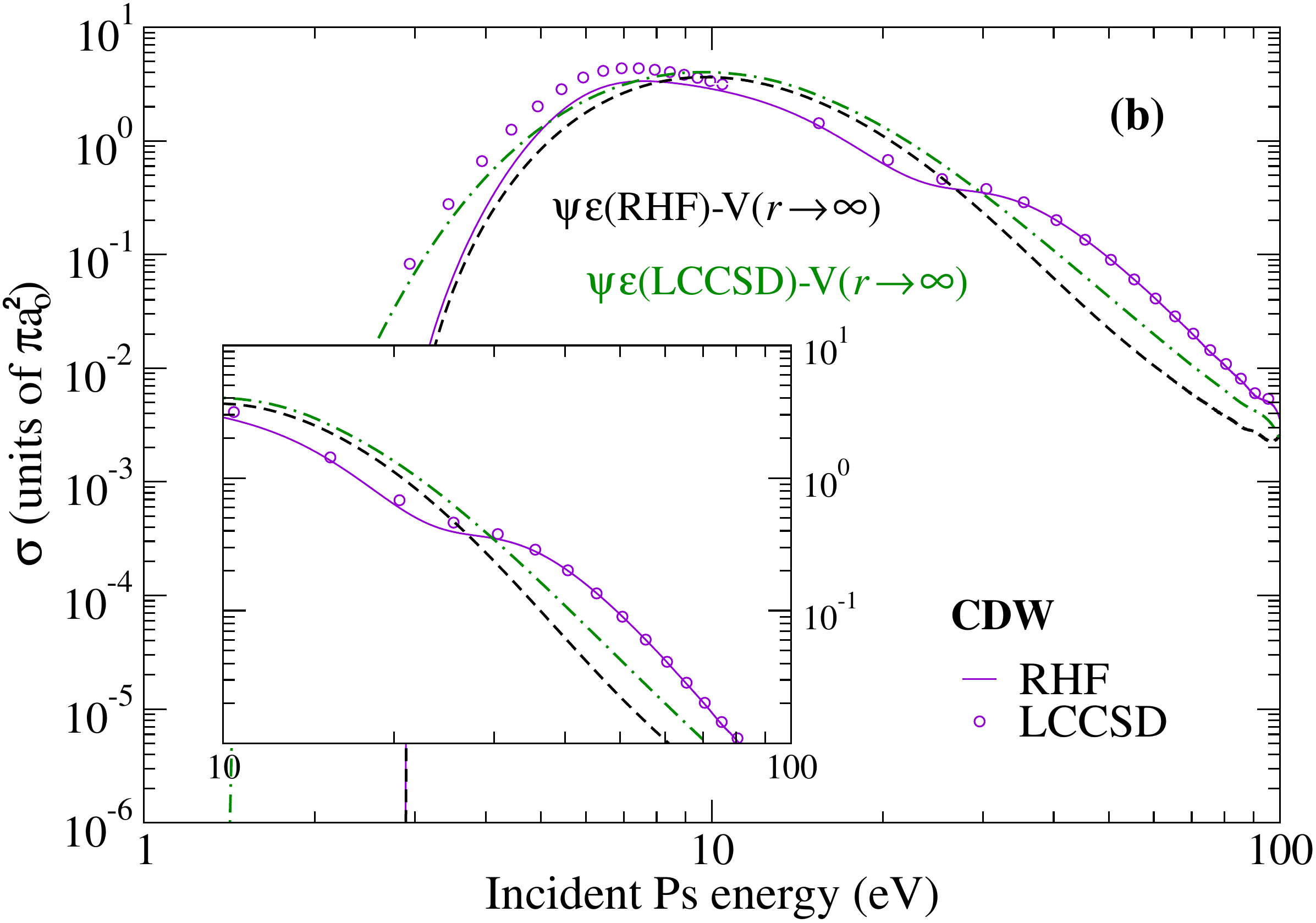}
\subcaption{\label{annexe-2}}
\end{subfigure}
\vskip-6mm
\caption{\textbf{(a)} Evolution of the Ps(${n=2}$) CBA partial cross sections as a function of the Ps impact energy. The AV5Z-MOD basis set is used to generate data associated with PsCl for: (dashed line) a combination of the SPP of the RHF radial wavefunction of the positron bound to Cl$^-$ with the LCCSD positron binding energy; (dashed-dotted line) a combination of the SPP of the LCCSD radial wavefunction of the positron bound to Cl$^-$ with the RHF positron binding energy. \textbf{(b)} Evolution of the Ps(${n=1}$) CDW partial cross sections as a function of the Ps impact energy using the AV5Z-MOD basis set to generate data associated with PsCl. The asymptotic forms of the potentials ${V(\boldsymbol{r})}$ given by the Eq. (\ref{es.19}) is considered (GBAR) in both approximations: (dashed line) RHF; (dashed-dotted line) LCCSD.}
\end{figure}
\vskip+1mm
The two figures presented above complete the discussion provided in the manuscript, in subsection \ref{subsection-4} and Appendix \ref{appendix-1} respectively.
\end{widetext}
\end{document}